\documentclass[preprint,authoryear,times,12pt]{elsarticle}
\usepackage[german,american]{babel}

\usepackage{latexsym}
\usepackage{ulem}

\usepackage{bm}
\usepackage{amsfonts}
\usepackage{amssymb}
\usepackage{amsmath}

\usepackage[OMLmathsfit]{isomath}

\def\msbi#1{\mathsfbfit{#1}}

\usepackage{color} % beamer
\definecolor{darkgreen}{rgb}{0,0.6,0}
\definecolor{gray}{rgb}{0.5,0.5,0.5}
\definecolor{mauve}{rgb}{0.58,0,0.82}

\graphicspath{{Figures/}}

\usepackage{float}
\floatstyle{ruled}
\newfloat{algorithmLD}{htb}{loa}
\floatname{algorithmLD}{Algorithm}
\newfloat{algorithmHM}{htb}{loa}
\floatname{algorithmHM}{Algorithm}
\usepackage{caption}

\usepackage{epic,eepic}

\journal{...}

\begin{document} 

\begin{frontmatter}

\title{On the higher-order pseudo-continuum characterization of 
discrete kinematic results from experimental measurement or 
discrete simulation} 

\author{Mohammad Khorrami${}^{1}$, 
Jaber R.~Mianroodi${}^{1}$, 
Bob Svendsen$^{1,2}$}
\address{${}^{1}$Microstructure Physics and Alloy Design,
Max-Planck-Institut f\"ur Eisenforschung GmbH, D-40237 D\"usseldorf, Germany\\
${}^{2}$Material Mechanics, RWTH Aachen, Aachen, D-54062, Germany}

\begin{abstract}

The purpose of this work is the development and determination of higher-order 
continuum-like kinematic measures which characterize discrete kinematic 
data obtained from experimental measurement (e.g., digital image correlation) 
or kinematic results from discrete modeling 
and simulation (e.g., molecular statics, molecular dynamics, or quantum DFT). 
From a continuum point of view, such data or results 
are in general non-affine and incompatible, due for example to 
shear banding, material defects, or microstructure. 
To characterize such information in a (pseudo-) continuum fashion, the concept of 
discrete local deformation is introduced and exploited. The corresponding 
measures are determined in a purely discrete fashion independent of 
any relation to continuum fields. 
Demonstration and verification of the approach is carried out with 
the help of example applications based on non-affine and incompatible 
displacement information. 
In particular, for the latter case, molecular statics results for the displacement 
of atoms in and around a dislocation core in fcc Au are employed. 
The corresponding characterization of lattice distortion in and around 
the core in terms of higher-order discrete local deformation measures 
clearly shows that, even in the simplest case of planar cores, 
such distortion is only partly characterized by the Nye tensor. 

\end{abstract}


\end{frontmatter}

\section{Introduction}
\label{sec:IntDL} 

Experimental characterization methods like digital image correlation 
\cite[DIC:~][]{Sutton2009} divide a "region of interest" (ROI) 
\(R\) on the specimen surface into a union \(R=\bigcup_{i}R_{i}\) 
(not necessarily disjoint) of subregions \(R_{i}\). 
In the case of local subset DIC, these subregions are assumed 
to deform affinely with respect to their center during loading. The 
corresponding DIC model for the global deformation field \(\bm{\chi}\) 
takes the form 
\(
\bm{\chi}_{\mathrm{DIC}}(\bm{x}_{\mathrm{r}})
:=\sum_{i}\kappa_{i}(\bm{x}_{\mathrm{r}})
\,\lbrack
\bm{x}_{i\mathrm{c}}
+\bm{F}_{\!i}(\bm{x}_{\mathrm{r}}-\bm{x}_{i\mathrm{r}})
\rbrack
\), 
where "r" refers to reference (e.g., initial), and "c" to current, configuration. 
Here, \(\kappa_{i}\) is the characteristic function of \(R_{i}\), 
\(\bm{x}_{i\mathrm{c}}\) its current center, and 
\(\bm{F}_{\!i}\) the corresponding affine local deformation. 
In the case of global DIC, global compatibility is imposed 
\cite[e.g.,][]{Lu2000,Pan2015}, e.g., via discretization based on 
the finite element method (FEM). Employing XFEM in this latter 
context, global DIC has also been employed to characterize 
discontinuous deformation due to cracks and shear bands 
\cite[e.g.,][]{Rethore2007,Rethore2008}. Most recently, augmented 
Lagrangian DIC \cite[ALDIC:~][]{Yang2019} has been introduced, 
based in particular on the auxiliary (compatible) deformation field 
\(
\bm{\chi}_{\mathrm{ALDIC}}(\bm{x}_{\mathrm{r}})
\) 
such that
\(
\bm{x}_{i\mathrm{c}}
=\bm{\chi}_{\mathrm{ALDIC}}(\bm{x}_{i\mathrm{r}})
\)
and
\(
\bm{F}_{\!i}
=\nabla_{\!\mathrm{r}}\bm{\chi}_{\mathrm{ALDIC}}(\bm{x}_{i\mathrm{r}})
\). 
Except in the case of XFEM-based DIC, then, one assumes from the 
start that specimen deformation is locally affine or compatible. 
Consequently, any information in the data on non-affine or incompatible 
local deformation (e.g., due to shear banding, defects, microstructure) 
is lost in the characterization. 

Another example of discrete displacement "data" is represented the 
displacement of atoms in a crystalline lattice subject to loading. In general, the 
displacement of atoms in the neighborhood of a given atom is neither affine 
nor directly related to the localization of a continuum deformation field. 
A classic example of this is atomic displacement in the neighborhood of 
atoms in a dislocation core, often characterized by measures such as the 
differential displacement 
\cite[e.g.,][]{Vitek1970,Duesbery1998} or the (geometrically linear) Nye 
tensor \cite[][]{Nye53}. Such measures are commonly employed 
to characterize corresponding atomistic or \textit{ab initio} results 
\cite[e.g.,][]{Rodney2017}. 

Under the assumption that the change in relative separation between 
a given atom and those in a certain (e.g., first nearest-) neighborhood 
of this atom is affine, \cite{Hartley2005a} and \cite{Shimizu2007} 
developed methods to determine corresponding first-order local (pseudo-) 
deformation measures for atomic neighborhoods from atomic position 
information. Apparently unaware of the work of \cite{Shimizu2007}, 
\cite{Gullet2008} developed a similar approach and applied it to 
the determination of pseudo-continuum finite strain measures from 
atomistic simulation data. 
Likewise, \cite{Zimmerman2009} developed an approach analogous to 
that of \cite{Shimizu2007} and extended it to second-order. They applied 
their approach to analyze the deformation fields for a one-dimensional 
atomic chain, a biaxially stretched thin film containing a surface ledge, 
and an fcc metal subject to nano-indentation. 
More recently, \cite{Tucker2011} employed the approaches of 
\cite{Shimizu2007} and \cite{Zimmerman2009} to formulate 
pseudo-continuum kinematic measures for results from molecular 
dynamics simulations. In contrast to these works, \cite{Zhang2015} 
fit continuum deformation fields and tractions to molecular dynamics 
results via weighted least-squares minimization. 

The purpose of the current work is to develop an approach capable 
of characterizing discrete displacement data or results which may 
contain information on local deformation which from a continuum 
point of view is non-affine or incompatible. In doing this, a 
higher-order kinematic characterization of atomic position information 
going beyond measures like the differential displacement and Nye tensor 
is obtained. To these ends, the concept of discrete local 
deformation (of order \(m\))\footnote{For example, \cite{Gullet2008} 
employ the term "discrete deformation gradient", which corresponds to a 
discrete local deformation of order one here. Applications of 
this concept are pursued in this work for both finite and infinite (periodic) regions. 
In the latter case, where no form or shape change of a finite region 
is involved, "distortion" would be more appropriate than "deformation". 
For simplicity, however, "deformation" is used in both cases.} 
is employed in this work. 
The discrete measures involved represent a generalization of those introduced by 
\cite{Hartley2005a}, \cite{Shimizu2007} and \cite{Zimmerman2009} 
for pseudo-continuum kinematic characterization of atomic 
displacement information. Since they are purely discrete in nature, 
these measures are independent of any interpretation or assumption 
concerning their possible relation to the deformation of a continuum. 
As such, they have the same character as the experimental data or 
atomisitic results on which they are based. 

%Besides being of intrinsic interest, determination of such 
%measures is relevant for comparison of atomistic results with predictions 
%from continuum modeling approaches like phase-field dislocation dynamics 
%\cite[e.g.,][]{Hun13,Xu2019} or atomistic phase field microelasticity 
%\cite[e.g.,][]{Mia15,Mianroodi2016a}. 

After a brief summary of required mathematical notation and results in 
Section \ref{sec:NotMatDL}, the work begins with a brief review of the 
concept of discrete local deformation of order \(m\) 
in Section \ref{sec:DefLocDet3D}. 
This is followed by the development of a method to determine discrete 
local deformation measures based on 3D position data. 
In the process, the first-order approaches 
of \cite{Hartley2005a} and \cite{Shimizu2007} are generalized to higher 
order. As an example application, atomic position configurations in the 
dislocation core of straight edge and screw dislocations in Au are employed 
in Section \ref{sec:CorDisAppDL} to determine first and second-order 
discrete local deformation measures of atomic neighborhoods. This is 
followed by the formulation and discussion of fields induced by discrete 
local dislocation measures in Section \ref{sec:DLDIntComPse}. After 
discussing the relation of the current treatment to selected previous work 
in Section \ref{sec:WorPreSel}, the current work is summarized in 
Section \ref{sec:DisSumDL}. This includes a discussion of further 
aspects and potential developments. Additional background and mathematical 
details are given in the appendix. 

\section{Mathematical preliminaries \& notation}
\label{sec:NotMatDL}

Let \(\mathbb{E}^{3}\) represent three-dimensional Euclidean point space 
with translation / vector space \(\mathbb{V}^{3}\). 
In this work, lower-case bold italic characters represent 
elements of \(\mathbb{E}^{3}\) or \(\mathbb{V}^{3}\). In particular, let 
\(\bm{i}_{1}=\bm{i}_{x}\), \(\bm{i}_{2}=\bm{i}_{y}\), and 
\(\bm{i}_{3}=\bm{i}_{z}\) represent the Cartesian basis vectors. 
Second-order tensors 
\(\bm{A},\bm{B},\ldots\in\mathrm{Lin}(\mathbb{V}^{3},\mathbb{V}^{3})\) 
are represented by upper-case bold italic characters, with \(\bm{I}\) 
the second-order identity. Let 
\(
\mathcal{A}\cdot\mathcal{B}
=\sum_{ijk\ldots}A_{ijk\ldots}B_{ijk\ldots}
\in\mathbb{R}
\) 
represent the scalar product of two arbitrary-order tensors \(\mathcal{A}\) 
and \(\mathcal{B}\). Given this product \(\bm{a}\cdot\bm{b}\) on vectors, 
for example, one can define the transpose \(\bm{A}^{\!\mathrm{T}}\) of any 
\(\bm{A}\) by 
\(
\bm{c}\cdot\bm{A}^{\!\mathrm{T}}\bm{b}:=\bm{b}\cdot\bm{A}\bm{c}
\) 
for any \(\bm{b},\bm{c}\). 
In turn, \(\bm{A}^{\!\mathrm{T}}\) determines the symmetric 
\(
\mathop{\mathrm{sym}}\bm{A}
:=\frac{1}{2}(\bm{A}+\bm{A}^{\!\mathrm{T}})
\) 
and skew-symmetric 
\(
\mathop{\mathrm{skw}}\bm{A}
:=\frac{1}{2}(\bm{A}-\bm{A}^{\!\mathrm{T}})
\) 
parts of any \(\bm{A}\). As usual, 
\(
\mathop{\mathrm{axv}}\bm{A}\times\bm{b}
:=\bm{A}\bm{b}
\) 
defines the axial vector \(\mathop{\mathrm{axv}}\bm{A}\) of any 
skew-symmetric \(\bm{A}\). Analogously, any vector \(\bm{a}\) induces 
a second-order ("axial") tensor \(\mathop{\mathrm{axt}}\bm{a}\) 
defined by \((\mathop{\mathrm{axt}}\bm{a})\,\bm{b}:=\bm{a}\times\bm{b}\). 
Note that 
\(
\mathop{\mathrm{axt}}\mathop{\mathrm{axv}}\bm{W}
=\bm{W}
\) 
and 
\(
\mathop{\mathrm{axv}}\mathop{\mathrm{axt}}\bm{a}
=\bm{a}
\). 
Note also that the axial tensor operation can be generalized to 
a second- or higher-order tensor \(\mathcal{A}\) via  
\(
((\mathop{\mathrm{axt}}\mathcal{A})\,\bm{a})\bm{b}
:=\mathcal{A}\,(\bm{a}\times\bm{b})
\), 
i.e., 
\(
(\mathop{\mathrm{axt}}\mathcal{A})\,\bm{a}
:=\mathcal{A}\mathop{\mathrm{axt}}\bm{a}
\). 

Let \(\mathrm{Lin}(\mathcal{X},\mathcal{Y})\) represent the set of all 
linear transformations between two linear (e.g., vector) spaces \(\mathcal{X}\) 
and \(\mathcal{Y}\). The concept of discrete local deformation employed 
in this work is based on the set 
\(
\mathrm{Lin}_{d}(\mathbb{V}^{3},\mathbb{V}^{3})
\cong
\mathrm{Lin}_{d+1}(\mathbb{V}^{3},\mathbb{R})
\) 
of all multilinear transformations of \(d\) vectors into a vector. 
Elements of \(\mathrm{Lin}_{d}(\mathbb{V}^{3},\mathbb{V}^{3})\) 
are symbolized by \(\msbi{A}^{(d)},\msbi{B}^{(d)},\ldots\) in what 
follows. 
In particular, 
\(
\msbi{A}^{(1)}
\in
\mathrm{Lin}_{1}(\mathbb{V}^{3},\mathbb{V}^{3})
=\mathrm{Lin}(\mathbb{V}^{3},\mathbb{V}^{3})
\) 
is then a second-order tensor. For any 
\(\msbi{A}^{(d)}\in\mathrm{Lin}_{d}(\mathbb{V}^{3},\mathbb{V}^{3})\), 
let 
\begin{equation}
\textstyle
\mathrm{sym}_{k}\msbi{A}^{(d)}
:=\frac{1}{k!}\sum_{\pi_{k}}
\msbi{A}_{\smash{\pi_{k}}}^{(d)}
\,,\quad
\mathrm{skw}_{k}\msbi{A}^{(d)}
:=\frac{1}{k!}\sum_{\pi_{k}}(-1)^{\pi_{k}}
\msbi{A}_{\smash{\pi_{k}}}^{(d)}
\,,\quad
k\leqslant d
\,,
\label{equ:SkwSymGen}
\end{equation}
represent its \(k\)-symmetric and \(k\)-skew-symmetric parts, respectively, 
where \(\msbi{A}_{\smash{\pi_{k}}}^{(d)}\) is a permutation of the last 
\(k\) arguments of \(\msbi{A}^{(d)}\). Of particular interest in the current 
work are the cases \(k=d\) and \(k=2\). In the latter case for example, 
\begin{equation}
\begin{array}{rcl}
((\mathrm{sym}_{2}\msbi{A}^{(d)})\bm{a})\bm{b}
&=&
\tfrac{1}{2}
(\msbi{A}^{(d)}\bm{a})\bm{b}
+\tfrac{1}{2}
(\msbi{A}^{(d)}\bm{b})\bm{a}
\,,\\ 
((\mathrm{skw}_{2}\msbi{A}^{(d)})\bm{a})\bm{b}
&=&
\tfrac{1}{2}
(\msbi{A}^{(d)}\bm{a})\bm{b}
-\tfrac{1}{2}
(\msbi{A}^{(d)}\bm{b})\bm{a}
\,.
\end{array}
\label{equ:SkwSymSwiDL}
\end{equation}
For \(k=2\), one can also define the axial "vector" 
\begin{equation}
(\mathrm{axv}_{2}\mathrm{skw}_{2}\msbi{A}^{(d)})
\,(\bm{a}\times\bm{b})
:=((\mathrm{skw}_{2}\msbi{A}^{(d)})\bm{a})\bm{b}
\label{equ:AxiSkwSwiDL}
\end{equation}
of \(\mathrm{skw}_{2}\msbi{A}^{(d)}\); then 
\(
(\mathrm{axv}_{2}\mathrm{skw}_{2}\msbi{A}^{(d)})
\mathop{\mathrm{axt}}\bm{a}
=(\mathrm{skw}_{2}\msbi{A}^{(d)})\bm{a}
\). 
Lastly, let 
\(
\mathrm{Sym}_{\smash{k,d}}(\mathbb{V}^{3},\mathbb{V}^{3})
\subset
\mathrm{Lin}_{d}(\mathbb{V}^{3},\mathbb{V}^{3})
\) 
with \(k\leqslant d\) represent the set of all 
\(
\msbi{A}^{(d)}
\in
\mathrm{Lin}_{d}(\mathbb{V}^{3},\mathbb{V}^{3})
\) 
for which 
\(
\mathrm{sym}_{k}\msbi{A}^{(d)}
=\msbi{A}^{(d)}
\) 
holds, i.e., the set of all \(k\)-symmetric elements of 
\(\mathrm{Lin}_{d}(\mathbb{V}^{3},\mathbb{V}^{3})\). 

Additional concepts and notation required in this work are introduced as 
we go along, or discussed in more detail in the appendix. 

\section{Determination of discrete local deformation from position data / results}
\label{sec:DefLocDet3D}

\subsection{Discrete local deformation of order one} 
\label{sec:OrdFirDL} 

Consider \(n\) points with time-dependent positions 
\(\bm{r}_{\!1}(t),\ldots,\bm{r}_{\!n}(t)\in\mathbb{E}^{3}\). 
Assume that \(\bm{r}_{1}(t),\ldots,\bm{r}_{n}(t)\) are known 
or have been determined for \(t\in\lbrace 0,t_{1},t_{2},\ldots\rbrace\). 
Let \(\msbi{F}_{\!\smash{\beta}+}^{(1)}(t)\) and 
\(\msbi{F}_{\!\smash{\beta}-}^{(1)}(t)\) be measures for 
discrete local deformation of order one associated with each 
\(\beta\in\lbrace 1,\ldots,n\rbrace\) for \(t>0\) such that
\begin{equation}
\begin{array}{rclcrcl}
\bm{s}_{\alpha\beta}(t)
&\approx&
\msbi{F}_{\!\smash{\beta+}}^{(1)}(t)\,\bm{s}_{\alpha\beta}(0)
\,,&&
\alpha
&\in&
N_{\smash{\beta}}^{(1)}(0)
\,,\\
\bm{s}_{\alpha\beta}(0)
&\approx&
\msbi{F}_{\!\smash{\beta-}}^{(1)}(t)\,\bm{s}_{\alpha\beta}(t)
\,,&&
\alpha
&\in&
N_{\smash{\beta}}^{(1)}(t)
\,,
\end{array}
\label{equ:VecDefSepDefLocOrdOne}
\end{equation} 
for \(t\in\lbrace t_{1},t_{2},\ldots\rbrace\). Here, 
\(\bm{s}_{\alpha\beta}:=\bm{r}_{\!\alpha}-\bm{r}_{\beta}\) 
represent the separation vector between \(\alpha\) and \(\beta\), and 
\(
N_{\smash{\beta}}^{(1)}
:=\lbrace 1_{\smash{\beta}}^{(1)},\ldots,n_{\smash{\beta}}^{(1)}\rbrace
\) 
is a set (list) of points in the neighborhood of \(\beta\). 
As evident in \eqref{equ:VecDefSepDefLocOrdOne}, 
\(\msbi{F}_{\!\smash{\beta+}}^{(1)}(t)\) is determined 
relative to \(\bm{r}_{1}(0),\ldots,\bm{r}_{n}(0)\), and 
\(\msbi{F}_{\!\smash{\beta-}}^{(1)}(t)\) relative to 
\(\bm{r}_{1}(t),\ldots,\bm{r}_{n}(t)\) for \(t>0\). The 
Cartesian component forms of \eqref{equ:VecDefSepDefLocOrdOne} 
determine the corresponding matrix forms 
\begin{equation}
\mathbf{S}_{\smash{\beta}}^{(1)}(t)
\approx
\mathbf{F}_{\!\smash{\beta+}}^{(1)}(t)
\,\mathbf{S}_{\smash{\beta}}^{(1)}(0)
\,,\quad
\mathbf{S}_{\smash{\beta}}^{(1)}(0)
\approx
\mathbf{F}_{\!\smash{\beta-}}^{(1)}(t)
\,\mathbf{S}_{\smash{\beta}}^{(1)}(t)
\,,
\label{equ:DefSepDefLocOrdOneMat}
\end{equation}
with 
\begin{equation}
\mathbf{S}_{\smash{\beta}}^{(d)}
:=\left\lbrack
\begin{array}{ccc}
\bm{i}_{1}\cdot\bm{s}_{1_{\smash{\beta}}^{(d)}\beta}
&\cdots
&\bm{i}_{1}\cdot\bm{s}_{\smash{n_{\smash{\beta}}^{(d)}\beta}}
\\
\bm{i}_{2}\cdot\bm{s}_{1_{\smash{\beta}}^{(d)}\beta}
&\cdots
&\bm{i}_{2}\cdot\bm{s}_{\smash{n_{\smash{\beta}}^{(d)}\beta}}
\\
\bm{i}_{3}\cdot\bm{s}_{1_{\smash{\beta}}^{(d)}\beta}
&\cdots
&\bm{i}_{3}\cdot\bm{s}_{\smash{n_{\smash{\beta}}^{(d)}\beta}}
\end{array}
\right\rbrack
\,,\quad
\mathbf{F}_{\!\smash{\beta}}^{(1)}
:=\left\lbrack
\begin{array}{ccc}
\lbrack\msbi{F}_{\!\smash{\beta}}^{(1)}\rbrack_{11}
&
\lbrack\msbi{F}_{\!\smash{\beta}}^{(1)}\rbrack_{12}
&
\lbrack\msbi{F}_{\!\smash{\beta}}^{(1)}\rbrack_{13}
\\
\lbrack\msbi{F}_{\!\smash{\beta}}^{(1)}\rbrack_{21}
&
\lbrack\msbi{F}_{\!\smash{\beta}}^{(1)}\rbrack_{22}
&
\lbrack\msbi{F}_{\!\smash{\beta}}^{(1)}\rbrack_{23}
\\
\lbrack\msbi{F}_{\!\smash{\beta}}^{(1)}\rbrack_{31}
&
\lbrack\msbi{F}_{\!\smash{\beta}}^{(1)}\rbrack_{32}
&
\lbrack\msbi{F}_{\!\smash{\beta}}^{(1)}\rbrack_{33}
\end{array}
\right\rbrack
\,,
\label{equ:DefLocOrdFirComMat}
\end{equation} 
\(n_{\smash{\beta}}^{(d)}:=|N_{\smash{\beta}}^{(d)}|\), 
and \(\lbrack\msbi{F}_{\!\smash{\beta}}^{(1)}\rbrack_{i\!j}
=\bm{i}_{i}\cdot\msbi{F}_{\!\smash{\beta}}^{(1)}\bm{i}_{\!j}\). 
For each \(\alpha\), \eqref{equ:VecDefSepDefLocOrdOne}$_{1,2}$ 
represent 3 equations in 9 unknowns; as such, 
\(\mathbf{F}_{\!\smash{\beta\pm}}^{(1)}(t)\) 
are overdetermined by \eqref{equ:VecDefSepDefLocOrdOne} for 
\(n_{\smash{\beta}}^{(1)}>3\). Following previous work 
\cite[e.g.,][]{Hartley2005a,Shimizu2007}, then, 
least-squares 
minimization\footnote{More generally, this should be based on 
weighted least-squares minimization as discussed by \cite{Gullet2008}; 
for simplicity, however, this is not done in the current work.} 
is employed for the fit of \(\msbi{F}_{\!\smash{\beta\pm}}^{(1)}\) 
to the data. As usual, the corresponding necessary conditions 
\begin{equation}
\begin{array}{rcl}
\mathbf{F}_{\!\smash{\beta+}}^{(1)}(t)
&=&
\lbrack
\mathbf{S}_{\smash{\beta}}^{(1)}(t)
\,\mathbf{S}_{\smash{\beta}}^{(1)\mathrm{T}}(0)
\rbrack
\,\lbrack
\mathbf{S}_{\smash{\beta}}^{(1)}(0)
\,\mathbf{S}_{\smash{\beta}}^{(1)\mathrm{T}}(0)
\rbrack^{-1}
\,,\\ 
\mathbf{F}_{\!\smash{\beta-}}^{(1)}(t)
&=&
\lbrack
\mathbf{S}_{\smash{\beta}}^{(1)}(0)
\,\mathbf{S}_{\smash{\beta}}^{(1)\mathrm{T}}(t)
\rbrack
\,\lbrack
\mathbf{S}_{\smash{\beta}}^{(1)}(t)
\,\mathbf{S}_{\smash{\beta}}^{(1)\mathrm{T}}(t)
\rbrack^{-1}
\,,
\end{array}
\label{equ:LagEulOrdOne}
\end{equation} 
determine \(\mathbf{F}_{\!\smash{\beta\pm}}^{(1)}(t)\) 
(i.e., in the least-squares sense). Since \eqref{equ:LagEulOrdOne} imply 
\begin{equation}
\mathbf{F}_{\!\smash{\beta-}}^{(1)}(t)
=\lbrack
\mathbf{S}_{\smash{\beta}}^{(1)}(0)
\,\mathbf{S}_{\smash{\beta}}^{(1)\mathrm{T}}(0)
\rbrack
\,\mathbf{F}_{\!\smash{\beta+}}^{(1)\mathrm{T}}(t)
\,\lbrack
\mathbf{S}_{\smash{\beta}}^{(1)}(t)
\,\mathbf{S}_{\smash{\beta}}^{(1)\mathrm{T}}(t)
\rbrack^{-1}
\,,
\end{equation} 
note that 
\(\msbi{F}_{\!\smash{\beta-}}^{(1)}\) and 
\(\msbi{F}_{\!\smash{\beta+}}^{(1)}\) are not inversely related 
in general. 
As discussed in more detail later, \(\mathbf{F}_{\!\smash{\beta-}}^{(1)}\) is 
considered by \cite{Hartley2005a}, and \(\mathbf{F}_{\!\smash{\beta+}}^{(1)}\) 
by \cite{Shimizu2007}. Both of these determine corresponding distortions 
\(
\msbi{H}_{\smash{\beta+}}^{(1)}(t)
:=\msbi{F}_{\!\smash{\beta+}}^{(1)}(t)-\bm{I}
\)
and 
\(
\msbi{H}_{\smash{\beta-}}^{(1)}(t)
:=\bm{I}-\msbi{F}_{\!\smash{\beta-}}^{(1)}(t)
\)
such that 
\(
\bm{u}_{\smash{\alpha\beta}}(t)
=\msbi{H}_{\smash{\beta+}}^{(1)}(t)
\,\bm{s}_{\alpha\beta}(0)
=\msbi{H}_{\smash{\beta-}}^{(1)}(t)
\,\bm{s}_{\alpha\beta}(t)
\) 
hold for the relative displacements 
\(
\bm{u}_{\smash{\alpha\beta}}(t)
:=\bm{u}_{\alpha}(t)-\bm{u}_{\beta}(t)
\) 
with \(\bm{u}_{\alpha}(t):=\bm{r}_{\alpha}(t)-\bm{r}_{\alpha}(0)\).

\subsection{Discrete local deformation of order two} 
\label{sec:OrdSecDL} 

Given \(\msbi{F}_{\!\smash{\beta\pm}}^{(1)}(t)\) as just determined for 
\(\beta\in\lbrace 1,\ldots,n\rbrace\) and \(t\in\lbrace 0,t_{1},t_{2},\ldots\rbrace\), 
\begin{equation}
\msbi{H}_{\smash{\alpha\beta\pm}}^{(1)}(t)
:=\msbi{F}_{\!\smash{\alpha\pm}}^{(1)}(t)
-\msbi{F}_{\!\smash{\beta\pm}}^{(1)}(t)
\label{equ:DefLocDisFirOrdDif}
\end{equation}
are known for \(\alpha,\beta\in\lbrace 1,\ldots,n\rbrace\) and 
\(t\in\lbrace 0,t_{1},t_{2},\ldots\rbrace\). 
Analogous to \eqref{equ:VecDefSepDefLocOrdOne}, then, assume there exists 
local deformation measures \(\msbi{F}_{\!\smash{\beta\pm}}^{(2)}(t)\) 
such that 
\begin{equation}
\begin{array}{rclcrcl}
\msbi{H}_{\smash{\alpha\beta+}}^{(1)}(t)
&\approx&
\msbi{F}_{\!\smash{\beta+}}^{(2)}(t)
\,\bm{s}_{\alpha\beta}(0)
\,,&&
\alpha
&\in&
N_{\smash{\beta}}^{(2)}(0)
\,,\\
\msbi{H}_{\smash{\alpha\beta-}}^{(1)}(t)
&\approx&
\msbi{F}_{\!\smash{\beta-}}^{(2)}(t)
\,\bm{s}_{\alpha\beta}(t)
\,,&&
\alpha
&\in&
N_{\smash{\beta}}^{(2)}(t)
\,,
\end{array}
\label{equ:VecDefSepDefLocTwoOrd}
\end{equation} 
for \(t\in\lbrace t_{1},t_{2},\ldots\rbrace\). In matrix form, 
\begin{equation}
\mathbf{H}_{\smash{\beta+}}^{(1)}(t)
\approx
\mathbf{F}_{\!\smash{\beta+}}^{(2)}(t)
\,\mathbf{S}_{\smash{\beta}}^{(2)}(0)
\,,\quad
\mathbf{H}_{\smash{\beta-}}^{(1)}(t)
\approx
\mathbf{F}_{\!\smash{\beta-}}^{(2)}(t)
\,\mathbf{S}_{\smash{\beta}}^{(2)}(t)
\,,
\label{equ:DefSepDefLocOrdTwoMat}
\end{equation}
analogous to \eqref{equ:DefSepDefLocOrdOneMat}, with 
\(\mathbf{S}_{\smash{\beta}}^{(2)}\) given by 
\eqref{equ:DefLocOrdFirComMat}${}_{1}$ for \(d=2\), 
\begin{equation}
\mathbf{H}_{\smash{\beta}}^{(1)}
:=\left\lbrack
\begin{array}{ccc}
\lbrack
\msbi{H}_{\smash{1_{\smash{\beta}}^{(2)}\beta}}^{(1)}
\rbrack_{11}
&\cdots
&\lbrack
\msbi{H}_{\smash{n_{\smash{\beta}}^{(2)}\beta}}^{(1)}
\rbrack_{11}
\\
\lbrack
\msbi{H}_{\smash{1_{\smash{\beta}}^{(2)}\beta}}^{(1)}
\rbrack_{12}
&\cdots
&\lbrack
\msbi{H}_{\smash{n_{\smash{\beta}}^{(2)}\beta}}^{(1)}
\rbrack_{12}
\\
\vdots
&\vdots
&\vdots
\\
\lbrack
\msbi{H}_{\smash{1_{\smash{\beta}}^{(2)}\beta}}^{(1)}
\rbrack_{32}
&\cdots
&\lbrack
\msbi{H}_{\smash{n_{\smash{\beta}}^{(2)}\beta}}^{(1)}
\rbrack_{32}
\\
\lbrack
\msbi{H}_{\smash{1_{\smash{\beta}}^{(2)}\beta}}^{(1)}
\rbrack_{33}
&\cdots
&\lbrack
\msbi{H}_{\smash{n_{\smash{\beta}}^{(2)}\beta}}^{(1)}
\rbrack_{33}
\end{array}
\right\rbrack
\,,\quad
\mathbf{F}_{\!\smash{\beta}}^{(2)}
:=\left\lbrack
\begin{array}{ccc}
\lbrack\msbi{F}_{\!\smash{\beta}}^{(2)}\rbrack_{111}
&\lbrack\msbi{F}_{\!\smash{\beta}}^{(2)}\rbrack_{112}
&\lbrack\msbi{F}_{\!\smash{\beta}}^{(2)}\rbrack_{113}
\\
\lbrack\msbi{F}_{\!\smash{\beta}}^{(2)}\rbrack_{121}
&\lbrack\msbi{F}_{\!\smash{\beta}}^{(2)}\rbrack_{122}
&\lbrack\msbi{F}_{\!\smash{\beta}}^{(2)}\rbrack_{123}
\\
\vdots
&\vdots
&\vdots
\\
\lbrack\msbi{F}_{\!\smash{\beta}}^{(2)}\rbrack_{321}
&\lbrack\msbi{F}_{\!\smash{\beta}}^{(2)}\rbrack_{322}
&\lbrack\msbi{F}_{\!\smash{\beta}}^{(2)}\rbrack_{323}
\\
\lbrack\msbi{F}_{\!\smash{\beta}}^{(2)}\rbrack_{331}
&\lbrack\msbi{F}_{\!\smash{\beta}}^{(2)}\rbrack_{332}
&\lbrack\msbi{F}_{\!\smash{\beta}}^{(2)}\rbrack_{333}
\end{array}
\right\rbrack
\,,
\end{equation} 
and \(\lbrack\msbi{F}_{\!\smash{\beta}}^{(2)}\rbrack_{i\!jk}
=\bm{i}_{i}\cdot(\msbi{F}_{\!\smash{\beta}}^{(2)}\bm{i}_{k})\bm{i}_{\!j}\). 
Analogous to \(\mathbf{F}_{\!\smash{\beta\pm}}^{(1)}(t)\) 
for \(n_{\smash{\beta}}^{(1)}>3^1\) in the order one case, 
note that \(\mathbf{F}_{\!\smash{\beta\pm}}^{(2)}(t)\) are 
overdetermined for \(n_{\smash{\beta}}^{(2)}>3^2\). 
Employing again least-squares minimization, one obtains 
\begin{equation}
\begin{array}{rcl}
\mathbf{F}_{\!\smash{\beta+}}^{(2)}(t)
&=&
\lbrack
\mathbf{H}_{\smash{\beta+}}^{(1)}(t)
\,\mathbf{S}_{\smash{\beta}}^{(2)\mathrm{T}}(0)
\rbrack
\,\lbrack
\mathbf{S}_{\smash{\beta}}^{(2)}(0)
\,\mathbf{S}_{\smash{\beta}}^{(2)\mathrm{T}}(0)
\rbrack^{-1}
\,,\\ 
\mathbf{F}_{\!\smash{\beta-}}^{(2)}(t)
&=&
\lbrack
\mathbf{H}_{\smash{\beta-}}^{(1)}(t)
\,\mathbf{S}_{\smash{\beta}}^{(2)\mathrm{T}}(t)
\rbrack
\,\lbrack
\mathbf{S}_{\smash{\beta}}^{(2)}(t)
\,\mathbf{S}_{\smash{\beta}}^{(2)\mathrm{T}}(t)
\rbrack^{-1}
\,,
\end{array}
\label{equ:LagEulTwoOrd}
\end{equation} 
analogous to \eqref{equ:LagEulOrdOne}. As 
evident from \eqref{equ:DefSepDefLocOrdTwoMat} or 
this last relation, in contrast to \(\msbi{F}_{\!\smash{\beta+}}^{(2)}\), 
the dependence of \(\msbi{F}_{\smash{\beta-}}^{(2)}\) on 
\(\bm{r}_{1}(0),\ldots,\bm{r}_{n}(0)\) is only implicit. 

\subsection{Discrete local deformation of higher order} 
\label{sec:OrdIDL} 

By analogy with the last subsection, 
given \(\msbi{F}_{\!\smash{\beta\pm}}^{(m-1)}(t)\) 
for \(\beta\in\lbrace 1,\ldots,n\rbrace\) 
and \(t\in\lbrace 0,t_{1},t_{2},\ldots\rbrace\), let 
\(\msbi{F}_{\!\smash{\beta\pm}}^{(m)}(t)\) be local deformation 
measures such that 
\begin{equation}
\begin{array}{rclcrcl}
\msbi{H}_{\smash{\alpha\beta+}}^{(m-1)}(t)
&\approx&
\msbi{F}_{\!\smash{\beta+}}^{(m)}(t)
\,\bm{s}_{\alpha\beta}(0)
\,,&&
\alpha
&\in&
N_{\smash{\beta}}^{(m)}(0)
\,,\\
\msbi{H}_{\smash{\alpha\beta-}}^{(m-1)}(t)
&\approx&
\msbi{F}_{\smash{\beta-}}^{(m)}(t)
\,\bm{s}_{\alpha\beta}(t)
\,,&&
\alpha
&\in&
N_{\smash{\beta}}^{(m)}(t)
\,,
\end{array}
\label{equ:VecDefSepDefLocOrdI}
\end{equation} 
for \(t\in\lbrace t_{1},t_{2},\ldots\rbrace\). As before, these can be expressed 
in matrix form 
\begin{equation}
\mathbf{H}_{\smash{\beta+}}^{(m-1)}(t)
\approx
\mathbf{F}_{\smash{\beta+}}^{(m)}(t)
\,\mathbf{S}_{\smash{\beta}}^{(m)}(0)
\,,\quad
\mathbf{H}_{\smash{\beta-}}^{(m-1)}(t)
\approx
\mathbf{F}_{\smash{\beta-}}^{(m)}(t)
\,\mathbf{S}_{\smash{\beta}}^{(m)}(t)
\,,
\label{equ:DefSepDefLocOrdIMat}
\end{equation} 
analogous to \eqref{equ:DefSepDefLocOrdTwoMat}, with 
\(\mathbf{S}_{\smash{\beta}}^{(m)}\) 
given by \eqref{equ:DefLocOrdFirComMat}${}_{1}$ for \(d=m\). Then 
\begin{equation}
\begin{array}{rcl}
\mathbf{F}_{\!\smash{\beta+}}^{(m)}(t)
&=&
\lbrack
\mathbf{H}_{\smash{\beta+}}^{(m-1)}(t)
\,\mathbf{S}_{\smash{\beta}}^{(m)\mathrm{T}}(0)
\rbrack
\,\lbrack
\mathbf{S}_{\smash{\beta}}^{(m)}(0)
\,\mathbf{S}_{\smash{\beta}}^{(m)\mathrm{T}}(0)
\rbrack^{-1}
\,,\\ 
\mathbf{F}_{\!\smash{\beta-}}^{(m)}(t)
&=&
\lbrack
\mathbf{H}_{\smash{\beta-}}^{(m-1)}(t)
\,\mathbf{S}_{\smash{\beta}}^{(m)\mathrm{T}}(t)
\rbrack
\,\lbrack
\mathbf{S}_{\smash{\beta}}^{(m)}(t)
\,\mathbf{S}_{\smash{\beta}}^{(m)\mathrm{T}}(t)
\rbrack^{-1}
\,,
\end{array}
\label{equ:LagEulOrdI}
\end{equation} 
follow via least-squares minimization, for 
\(n_{\smash{\beta}}^{(m)}\geqslant 3^{m}\), 
representing direct generalizations of \eqref{equ:LagEulTwoOrd}. 
Like in the order two case, 
\(\msbi{F}_{\smash{\beta-}}^{(m)}\) is only implicitly dependent 
on \(\bm{r}_{1}(0),\ldots,\bm{r}_{n}(0)\), 
in constrast to \(\msbi{F}_{\!\smash{\beta+}}^{(m)}\). 

\subsection{Discussion} 

As evident in \eqref{equ:LagEulOrdOne}, \eqref{equ:LagEulTwoOrd} and 
\eqref{equ:LagEulOrdI}, the constraints 
\begin{equation}
n_{\smash{\beta}}^{(d)}\geqslant 3^{d}
\,,\quad 
\det
\lbrack
\mathbf{S}_{\smash{\beta}}^{(d)}
\mathbf{S}_{\smash{\beta}}^{(d)\mathrm{T}}
\rbrack
>0
\,,
\label{equ:DefLocDetCon}
\end{equation} 
apply to the determination of \(\mathbf{F}_{\!\smash{\beta\pm}}^{(d)}\). 
Related to \eqref{equ:DefLocDetCon}${}_{2}$ is the constraint 
\begin{equation}
\begin{array}{rclcl}
\det\mathbf{F}_{\!\smash{\beta+}}^{(1)}(t)
&=&
\det
\,\lbrack
\mathbf{S}_{\smash{\beta}}^{(1)}(t)
\,\mathbf{S}_{\smash{\beta}}^{(1)\mathrm{T}}(0)
\rbrack
\,/\det
\,\lbrack
\mathbf{S}_{\smash{\beta}}^{(1)}(0)
\,\mathbf{S}_{\smash{\beta}}^{(1)\mathrm{T}}(0)
\rbrack
&>&
0
\,,\\ 
\det
\mathbf{F}_{\!\smash{\beta-}}^{(1)}(t)
&=&
\det
\,\lbrack
\mathbf{S}_{\smash{\beta}}^{(1)}(0)
\,\mathbf{S}_{\smash{\beta}}^{(1)\mathrm{T}}(t)
\rbrack
\,/\det
\,\lbrack
\mathbf{S}_{\smash{\beta}}^{(1)}(t)
\,\mathbf{S}_{\smash{\beta}}^{(1)\mathrm{T}}(t)
\rbrack
&>&
0
\,,
\end{array}
\label{equ:LagEulOrdOneDet}
\end{equation} 
for the invertibility of \(\mathbf{F}_{\!\smash{\beta\pm}}^{(1)}\) 
from \eqref{equ:LagEulOrdOne}. 
Again, \eqref{equ:DefLocDetCon}${}_{1}$ is a constraint on 
the minimum number of neighborhood points below 
which \(\mathbf{F}_{\!\smash{\beta\pm}}^{(d)}\) is not 
determinable. 

The above approach to determine discrete local deformation of order \(m\) 
\begin{equation}
(\msbi{F}^{(1)},\ldots,\msbi{F}^{(m)})
\,,\quad
\msbi{F}^{(d)}
\in
\mathrm{Lin}_{d}(\mathbb{V}^{3},\mathbb{V}^{3})
\,,\ \ 
d=1,\ldots,m
\,,\quad 
\det\msbi{F}^{(1)}
>0
\,,
\label{equ:DefLocDis}
\end{equation}
from discrete position information \(\bm{r}_{1},\ldots,\bm{r}_{n}\) 
is clearly completely independent of the source or physical nature 
of this information. As such, it can be applied equally well to results 
for position / displacement coming from 
(i) observation / measurement (e.g., DIC) or 
(ii) discrete modeling and simulation methods like molecular statics 
or dynamics. Examples of both of these cases are considered in the sequel. 

\section{Example applications} 
\label{sec:CorDisAppDL}

For simplicity, attention is restricted to the case that 
\(\bm{r}_{\!1}(0),\ldots,\bm{r}_{\!n}(0)\) are (perfect) lattice vectors 
in a cubic lattice. Then 
\(
\bm{r}_{\!\alpha}(0)
=r_{\mathrm{N}1}\,m_{\alpha k}\,\bm{i}_{k}
\) 
holds (summation convention) for all \(\alpha=1,\ldots,n\) with 
\(
m_{\alpha k}\in\mathbb{Z}
\) 
and \(r_{\mathrm{N}1}\) the (first) nearest-neighbor distance. Then 
\(
\bm{s}_{\alpha\beta}(0)
=r_{\mathrm{N}1}\,m_{\alpha\beta k}\,\bm{i}_{k}
\) 
with 
\(
m_{\alpha\beta k}:=m_{\alpha k}-m_{\beta k}\in\mathbb{Z}
\).  
For example, a regular grid of points represents a simple cubic lattice, and 
\(r_{\mathrm{N}1}\) is equal to the lattice constant \(a_{0}\). 
Points in neighborhoods of any \(\beta\) are then located in the interior 
or on the boundary of spheres with radii 
\(r_{\smash{\mathrm{N}1}}
<r_{\smash{\mathrm{N}2}}
<r_{\smash{\mathrm{N}3}}<\cdots
\) 
centered at \(\beta\); 
let \(S_{\!1\beta}\subset S_{\!2\beta}\subset S_{\!3\beta}\subset\cdots\) 
represent the corresponding lists of points. 
If for example \(\alpha\in S_{\!s\beta}(0)\), then clearly 
\(
|\bm{s}_{\alpha\beta}(0)|
=\sqrt{m_{\alpha\beta k}m_{\alpha\beta k}}
\ r_{\mathrm{N}1}
\leqslant 
r_{\smash{\mathrm{N}s}}
\) 
with 
\(
m_{\alpha\beta k}m_{\alpha\beta k}\in\mathbb{N}
\). 
In the simple cubic case, \(s=m_{\alpha\beta k}m_{\alpha\beta k}\)
implies \(\alpha\in S_{\!s\beta}(0)\). 

All results to follow are based on two discrete position configurations, i.e., 
the initial \(\bm{r}_{1}(0),\ldots,\bm{r}_{n}(0)\) and current 
(i.e., time \(t\)) or final \(\bm{r}_{1}(t),\ldots,\bm{r}_{n}(t)\) 
ones. In addition, to simplify the notation, define 
\(\bm{r}_{\smash{\alpha\mathrm{r}}}:=\bm{r}_{\alpha}(0)\) 
(subscript r for "reference") and 
\(\bm{r}_{\smash{\alpha\mathrm{c}}}:=\bm{r}_{\alpha}(t)\) 
(subscript c for "current").

\subsection{Displacement of points in a finite grid} 
\label{sec:ExpValDL}

Consider the deformation of a material containing a finite "grid" 
of "nodes" or points with positions \(\bm{r}_{1},\ldots,\bm{r}_{n}\). 
For example, these could be measurement points embedded in a (transparent) 
material which move with the material when it is loaded. In the case of 
subset-based local DIC for example \cite[e.g.][]{Pan2015}, 
these could be the subset centers in the (always finite) region of interest 
(ROI) of the specimen. In particular, let 
\(\bm{r}_{\smash{1\mathrm{r}}},\ldots,\bm{r}_{\smash{n\mathrm{r}}}\) 
correspond to a regular grid / simple cubic lattice with uniform 
spacing with grid spacing / lattice constant \(a_{0}\). 
Then \(\bm{r}_{\alpha}=\sum_{k=1}^{3}r_{\alpha k}\,\bm{i}_{k}\), 
\(r_{\smash{\alpha\mathrm{r}\,k}}\in a_{0}\lbrace 0,1,\ldots,n_{k}-1\rbrace\), 
and \(n=\prod_{k=1}^{3}n_{k}\). 

In what follows, let \(N_{\smash{\beta}}^{(d)}=S_{\!\smash{s\beta}}\), with 
\(S_{\!\smash{s\beta}}\) the smallest list of neighborhood points 
(i.e., smallest \(s\)) satisfying \(|S_{\!\smash{s\beta}}|\geqslant 3^{d}\) 
(i.e., \eqref{equ:DefLocDetCon}${}_{1}$). 
For a fixed, \textit{finite} regular grid or simple cubic lattice, we have 
4 types of points, i.e., (i) corner, (ii) edge, (ii) face, and (iv) interior. 
The results in this subsection are based on (i) 
\(N_{\smash{\beta}}^{(1)}=S_{\!\smash{1\beta}}\) (3 points), 
\(N_{\smash{\beta}}^{(2)}=S_{\!\smash{4\beta}}\) (10 points), 
\(N_{\smash{\beta}}^{(3)}=S_{\!\smash{9\beta}}\) (28 points), 
for corner points, (ii) 
\(N_{\smash{\beta}}^{(1)}=S_{\!\smash{1\beta}}\) (4 points),
\(N_{\smash{\beta}}^{(2)}=S_{\!\smash{2\beta}}\) (9 points), 
\(N_{\smash{\beta}}^{(3)}=S_{\!\smash{7\beta}}\) (27 points),
for edge points, (iii) 
\(N_{\smash{\beta}}^{(1)}=S_{\!\smash{1\beta}}\) (5 points),
\(N_{\smash{\beta}}^{(2)}=S_{\!\smash{2\beta}}\) (13 points), 
\(N_{\smash{\beta}}^{(3)}=S_{\!\smash{5\beta}}\) (30 points),
for face points, and (iv)  
\(N_{\smash{\beta}}^{(1)}=S_{\!\smash{1\beta}}\) (6 points),
\(N_{\smash{\beta}}^{(2)}=S_{\!\smash{2\beta}}\) (18 points), 
\(N_{\smash{\beta}}^{(3)}=S_{\!\smash{4\beta}}\) (29 points), 
for interior points. 

\begin{figure}[H]
\centering
\includegraphics[width=0.35\textwidth]{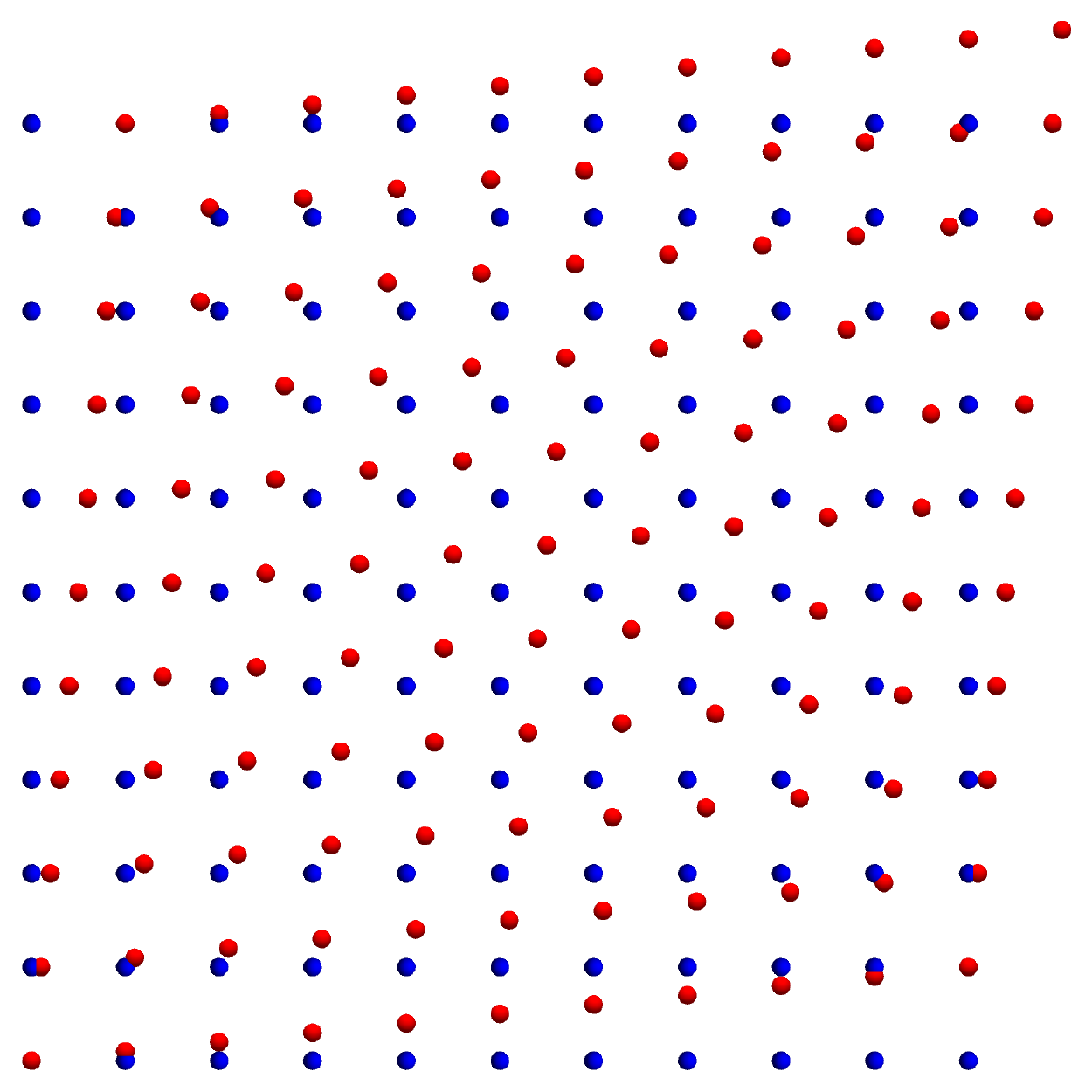}
\hspace{4mm}
\includegraphics[width=0.35\textwidth]{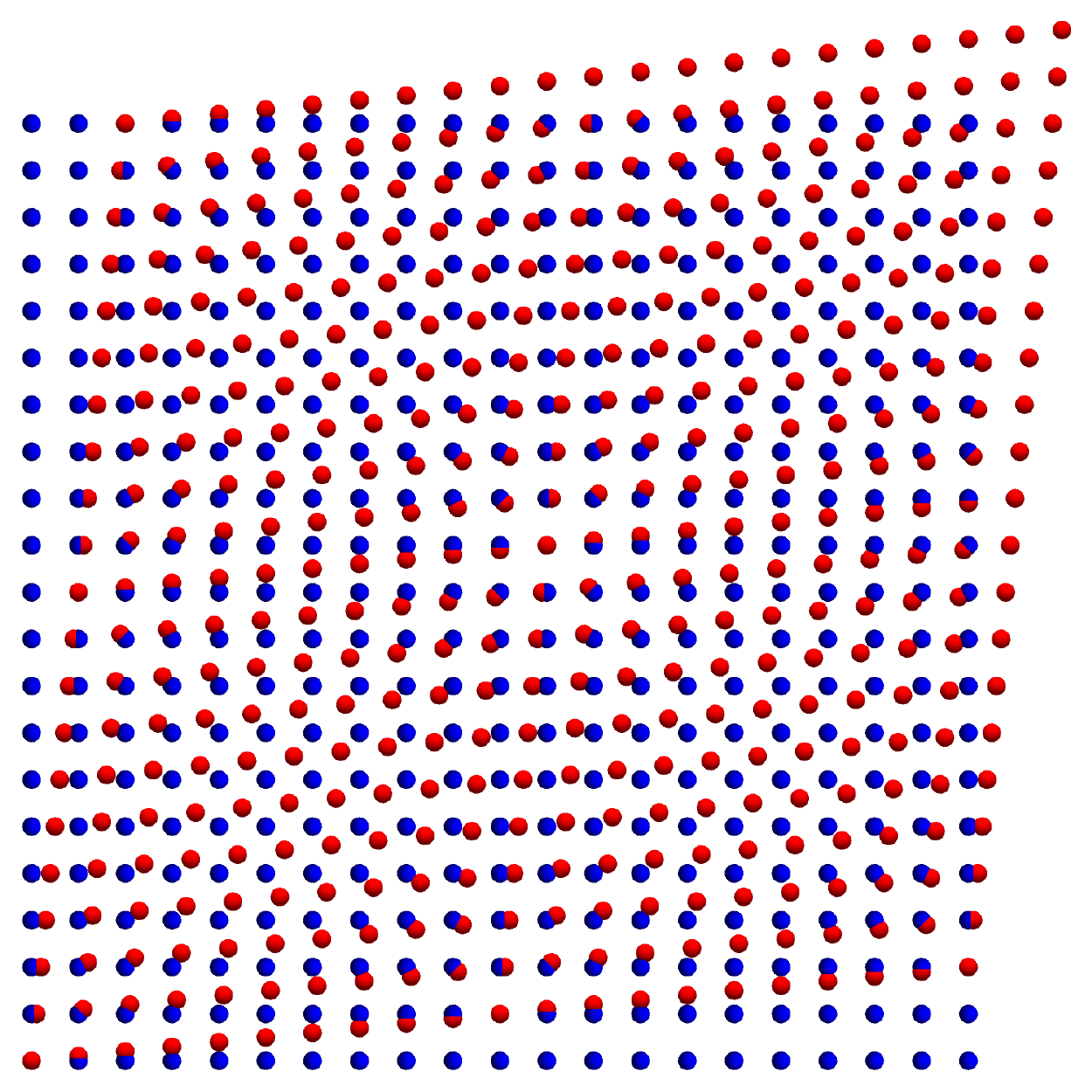}
\caption{Affine double (pure) shear of two point grids based 
on \eqref{equ:DefSheDouAff} 
for \(\varsigma_{1,2}=1\) and \(h_{1,2}=a_{0}(n_{1,2}-1)\). 
Left:~\(n=121\) points (\(n_{1,2}=11\)), \(a_{0}=1\). 
Right:~\(n=441\) points (\(n_{1,2}=21\)), \(a_{0}=\frac{1}{2}\). 
Blue:~initial grid. Red:~displaced grid.}
\label{fig:PoiShePur12}
\end{figure}

Consider first the (trivial) case of affine double shear 
\begin{equation}
\bm{\chi}(\bm{x}_{\mathrm{r}})
=\bm{x}_{\mathrm{r}}
+\varsigma_{1}\,(\bm{i}_{2}\cdot\bm{x}_{\mathrm{r}}/h_{2})\,\bm{i}_{1}
+\varsigma_{2}\,(\bm{i}_{1}\cdot\bm{x}_{\mathrm{r}}/h_{1})\,\bm{i}_{2}
\,.
\label{equ:DefSheDouAff}
\end{equation}
This is visualized in Figure \ref{fig:PoiShePur12}. 
From \eqref{equ:DefSheDouAff} follow 
\(
\nabla_{\!\smash{\mathrm{r}}}^{(1)}\bm{\chi}(\bm{x}_{\mathrm{r}})
=\bm{I}
+(\varsigma_{1}/h_{2})\,\bm{i}_{1}\otimes\bm{i}_{2}
+(\varsigma_{2}/h_{1})\,\bm{i}_{2}\otimes\bm{i}_{1}
\) 
and 
\(
\nabla_{\!\smash{\mathrm{r}}}^{(d)}\bm{\chi}(\bm{x}_{\mathrm{r}})
=\bm{0}
\) 
for 
\(d\geqslant 2\). 
A fit of \(\msbi{F}_{\!\smash{\beta\pm}}^{(1)}\) to the data 
in Figure \ref{fig:PoiShePur12} results in fit errors  
\(
|\msbi{F}_{\!\smash{\beta+}}^{(1)}
-\nabla_{\!\smash{\mathrm{r}}}^{(1)}\bm{\chi}(\bm{r}_{\beta\mathrm{r}})|
\)
and
\(
|\msbi{F}_{\!\smash{\beta-}}^{(1)}
-\nabla_{\!\smash{\mathrm{c}}}^{(1)}
\bm{\chi}^{-1}(\bm{r}_{\beta\mathrm{c}})|
\)
of machine precision. 
In addition, the position "data" in Figure \ref{fig:PoiShePur12} determine 
\(\msbi{F}_{\!\smash{\beta\pm}}^{(d)}=\bm{0}\) for \(d>1\) to machine 
precision at all \(\beta\). 
Consequently, determination of \(\msbi{F}_{\!\smash{\beta\pm}}^{(1)}\) 
recovers the theoretical result independent of resolution in the affine case. 

\begin{figure}[H]
\centering
\includegraphics[width=0.4\textwidth]{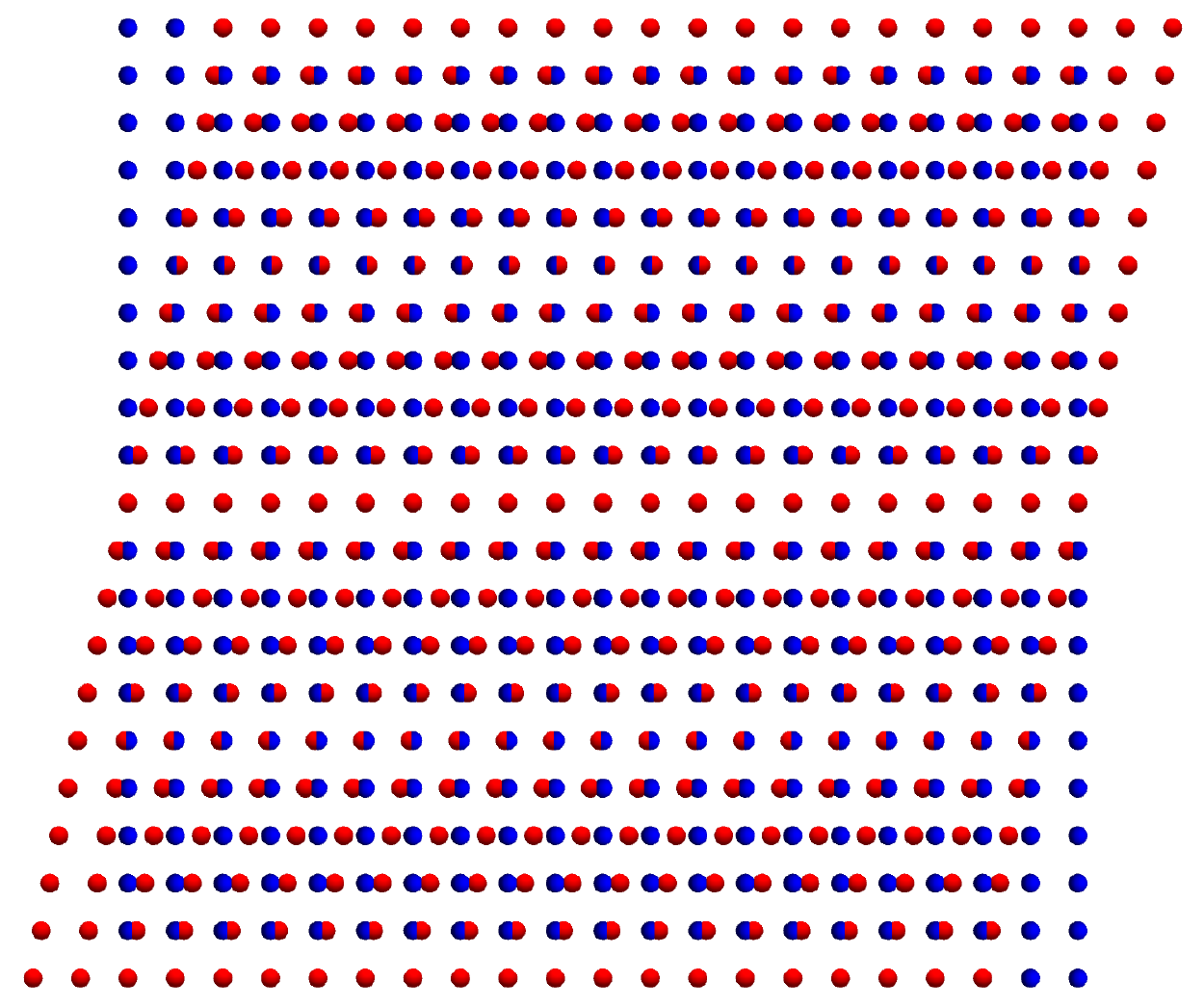}
\hspace{5mm}
\includegraphics[width=0.4\textwidth]{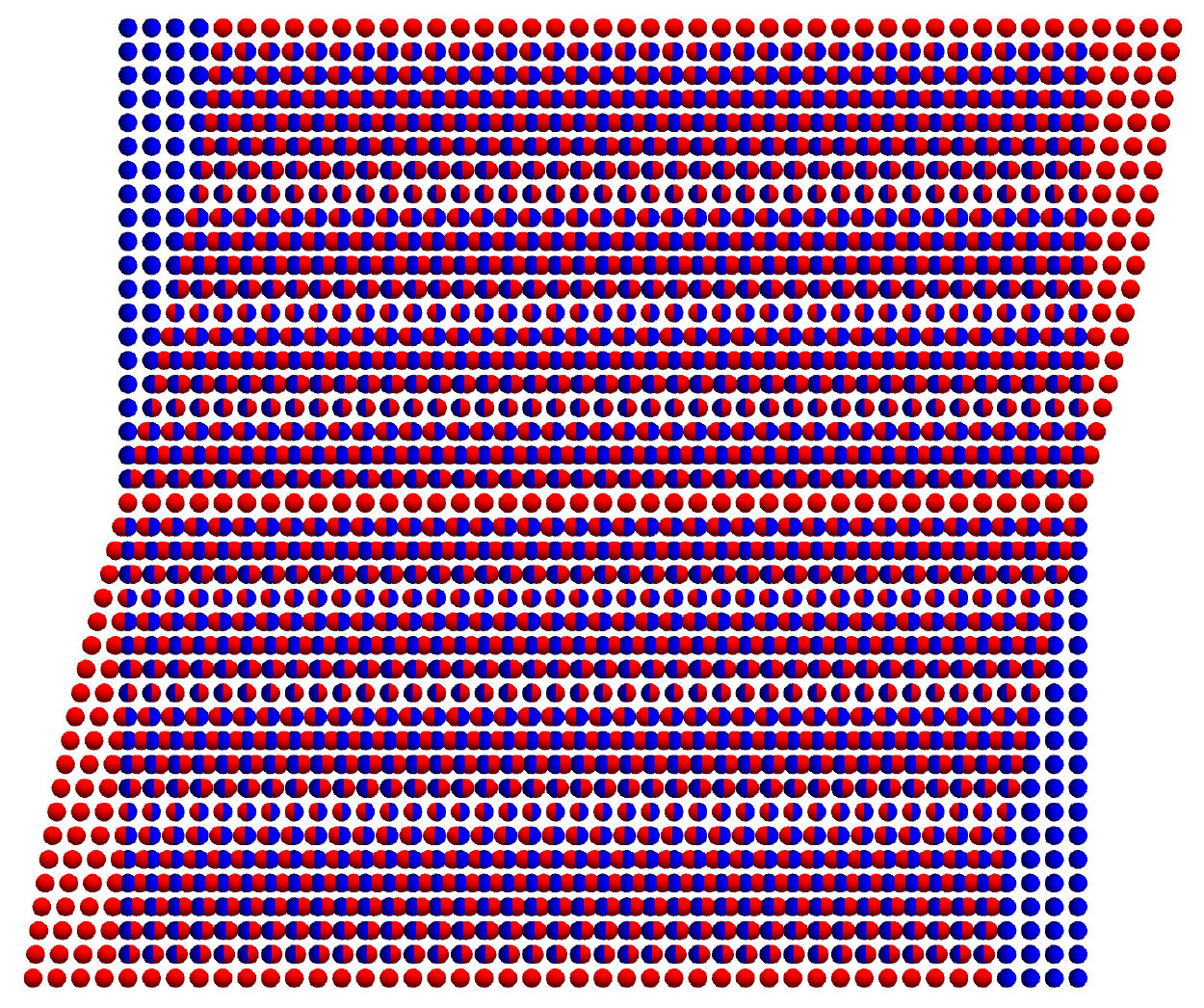}
\caption{Non-affine shear of two point grids based on \eqref{equ:DefSheAffNon} 
for \(\varsigma_{1}=2\) and \(c_{2}=5\). 
Left:~\(n=441\) points (\(n_{1,2}=21\)), \(a_{0}=\frac{1}{2}\), \(h_{2}=10\). 
Right:~\(n=1681\) points (\(n_{1,2}=41\)), \(a_{0}=\frac{1}{4}\), \(h_{2}=5\). 
Blue:~initial grid. Red:~displaced grid.}
\label{fig:PoiSheAffNon12}
\end{figure}

A less trivial case is represented by the non-affine shear
\begin{equation}
\bm{\chi}(\bm{x}_{\mathrm{r}})
=\bm{x}_{\mathrm{r}}
+\varsigma_{1}f((\bm{i}_{2}\cdot\bm{x}_{\mathrm{r}}-c_{2})/h_{2})
\,\bm{i}_{1}
\,,\quad 
f(x):=\frac{\tanh(x)}{2\tanh(c_{2}/h_{2})}
\,.
\label{equ:DefSheAffNon}
\end{equation}
This is displayed for two different values of 
\(h_{2}\) in Figure \ref{fig:PoiSheAffNon12}. 
In this case, we have 
\begin{equation}
\begin{array}{rcl}
\nabla_{\!\smash{\mathrm{r}}}^{(1)}\bm{\chi}(\bm{x}_{\mathrm{r}})
&=&
\bm{I}
+(\varsigma_{1}/h_{2})
\,f^{\prime}
((\bm{i}_{2}\cdot\bm{x}_{\mathrm{r}}-c_{2})/h_{2})
\,\bm{i}_{1}\otimes\bm{i}_{2}
\,,\\
\nabla_{\!\smash{\mathrm{r}}}^{(2)}\bm{\chi}(\bm{x}_{\mathrm{r}})
&=&
(\varsigma_{1}/h_{\smash{2}}^{2})
\,f^{\prime\prime}
((\bm{i}_{2}\cdot\bm{x}_{\mathrm{r}}-c_{2})/h_{2})
\,\bm{i}_{1}\otimes\bm{i}_{2}\otimes\bm{i}_{2}
\,,\\
\nabla_{\!\smash{\mathrm{r}}}^{(3)}\bm{\chi}(\bm{x}_{\mathrm{r}})
&=&
(\varsigma_{1}/h_{\smash{2}}^{3})
\,f^{\prime\prime\prime}
((\bm{i}_{2}\cdot\bm{x}_{\mathrm{r}}-c_{2})/h_{2})
\,\bm{i}_{1}\otimes\bm{i}_{2}\otimes\bm{i}_{2}\otimes\bm{i}_{2}
\,,\\
&\vdots&
\end{array}
\label{equ:GraDefSheAffNon}
\end{equation}
from \eqref{equ:DefSheAffNon}, where 
\(\nabla^{(d)}:=\nabla\circ\cdots\circ\nabla\) (\(d\) times). 
As shown in Figure \ref{fig:PoiSheAffNon12}, 
\(h_{2}\) controls the "amount" or "degree" of non-affinity. 
Indeed, for "large" \(h_{2}\), \(\tanh(x)\) is well-approximated by 
\(\tanh(x)\approx x\), and \eqref{equ:DefSheAffNon} is 
nearly affine. As \(h_{2}\) decreases, the non-linear terms in 
\(\tanh(x)\) become significant, and the non-affinity of 
\eqref{equ:DefSheAffNon} increases. 

\begin{figure}[H]
\centering
\includegraphics[width=0.4\textwidth]{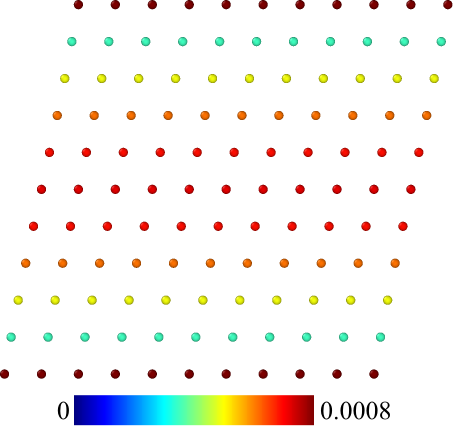}
\hspace{5mm}
\includegraphics[width=0.4\textwidth]{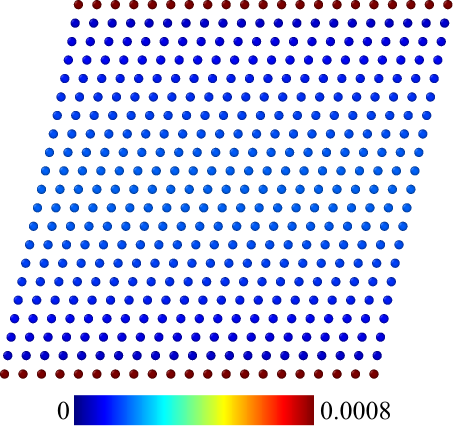}
\caption{Error \(|\lbrack\msbi{F}_{\smash{\beta+}}^{(1)}\rbrack_{12}
-(\varsigma_{1}/h_{2})\,f^{\prime}
((\bm{i}_{2}\cdot\bm{r}_{\smash{\beta\mathrm{r}}}-c_{2})/h_{2})|\) 
for \(h_{2}=10\) and two resolutions.\newline 
Left:~\(a_{0}=1\), \(n_{1,2}=11\). 
Right:~\(a_{0}=\frac{1}{2}\), \(n_{1,2}=21\). 
}
\label{fig:PoiDisSheBiPlu112One}
\end{figure}

In the following, results are presented for the accuracy of the 
largest component of \(\msbi{F}_{\!\smash{\beta+}}^{(d)}\) 
(\(d=1,2,3\)) assuming \(\varsigma_{1}=2\) and \(c_{2}=5\) as in 
Figure \ref{fig:PoiSheAffNon12}. To begin, consider these results for 
\(\lbrack\msbi{F}_{\smash{\beta+}}^{(1)}\rbrack_{12}\) with 
\(h_{2}=10\) for two resolutions in Figure \ref{fig:PoiDisSheBiPlu112One}. 
Note that the maximum error of 0.08\% is at the upper and lower 
boundaries of the region normal to the direction \(\pm\bm{i}_{2}\) 
of change in shear. As expected, there is an increase in accuracy with 
increasing resolution as documented in Figure \ref{fig:PoiDisSheBiPlu112One}. 

\begin{figure}[H]
\centering
\includegraphics[width=0.4\textwidth]{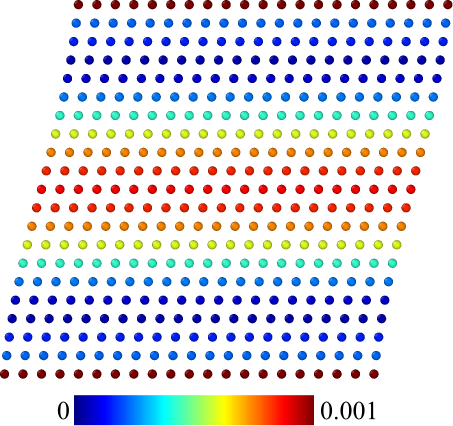}
\hspace{5mm}
\includegraphics[width=0.4\textwidth]{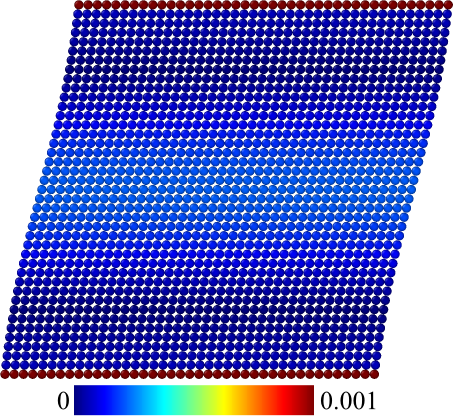}
\caption{Error \(|\lbrack\msbi{F}_{\smash{\beta+}}^{(1)}\rbrack_{12}
-(\varsigma_{1}/h_{2})\,f^{\prime}
((\bm{i}_{2}\cdot\bm{r}_{\smash{\beta\mathrm{r}}}-c_{2})/h_{2})|\) 
for \(h_{2}=5\) and two resolutions.\newline  
Left:~\(a_{0}=\frac{1}{2}\), \(n_{1,2}=21\).
Right:~\(a_{0}=\frac{1}{4}\), \(n_{1,2}=41\). 
}
\label{fig:PoiDisSheBiPlu112Two}
\end{figure}

These results and trends also hold for increasing non-affinity, as shown 
by the results in Figure \ref{fig:PoiDisSheBiPlu112Two} for \(h_{2}=5\).  
Comparison of the results in Figure \ref{fig:PoiDisSheBiPlu112Two} (left) 
with those in Figure \ref{fig:PoiDisSheBiPlu112One} (right) documents the 
expected increase in the error of 
\(\lbrack\msbi{F}_{\smash{\beta+}}^{(1)}\rbrack_{12}\)
as \(h_{2}\) decreases (i.e., as the "amount" of non-affinity increases) 
at a fixed resolution. 

\begin{figure}[H]
\centering
\includegraphics[width=0.4\textwidth]{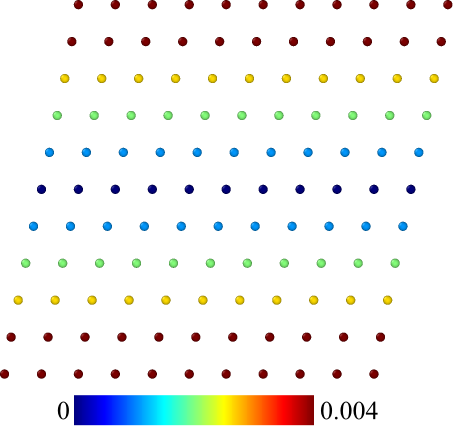}
\hspace{5mm}
\includegraphics[width=0.4\textwidth]{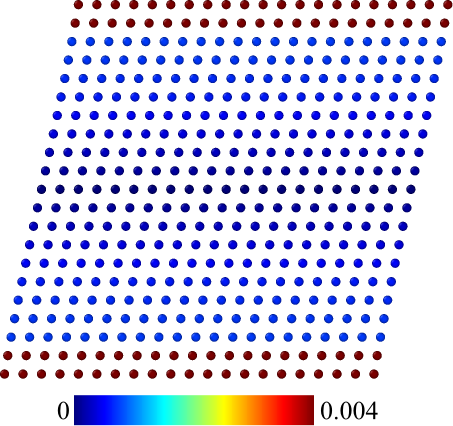}
\caption{Error \(|h_{2}\lbrack\msbi{F}_{\smash{\beta+}}^{(2)}\rbrack_{122}
-(\varsigma_{1}/h_{2})
\,f^{\prime\prime}
((\bm{i}_{2}\cdot\bm{r}_{\smash{\beta\mathrm{r}}}-c_{2})/h_{2})|\) 
for \(h_{2}=10\) and two resolutions.\newline  
Left:~\(a_{0}=1\), \(n_{1,2}=11\). 
Right:~\(a_{0}=\frac{1}{2}\), \(n_{1,2}=21\). 
}
\label{fig:PoiDisSheBiPlu2122One}
\end{figure}

\begin{figure}[H]
\centering
\includegraphics[width=0.4\textwidth]{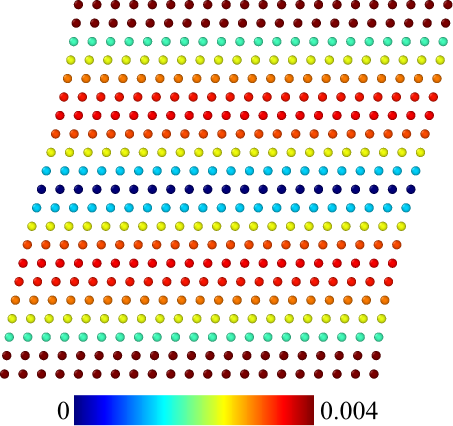}
\hspace{5mm}
\includegraphics[width=0.4\textwidth]{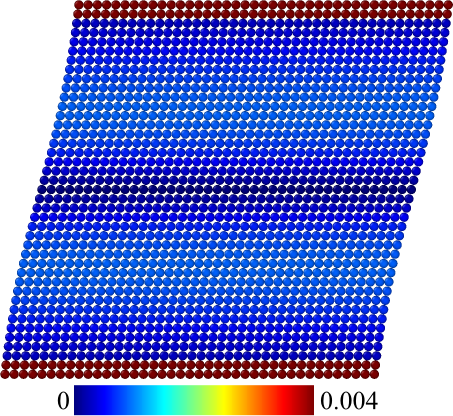}
\caption{Error \(|h_{2}\lbrack\msbi{F}_{\smash{\beta+}}^{(2)}\rbrack_{122}
-(\varsigma_{1}/h_{2})
\,f^{\prime\prime}
((\bm{i}_{2}\cdot\bm{r}_{\smash{\beta\mathrm{r}}}-c_{2})/h_{2})|\) 
for \(h_{2}=5\) and two resolutions.\newline  
Left:~\(a_{0}=\frac{1}{2}\), \(n_{1,2}=21\).
Right:~\(a_{0}=\frac{1}{4}\), \(n_{1,2}=41\). 
}
\label{fig:PoiDisSheBiPlu2122Two}
\end{figure}

Analogous results for the largest component 
\(\lbrack\msbi{F}_{\smash{\beta+}}^{(2)}\rbrack_{122}\) of 
\(\msbi{F}_{\!\smash{\beta+}}^{(2)}\) 
are shown in Figure \ref{fig:PoiDisSheBiPlu2122One} for \(h_{2}=10\)
and in Figure \ref{fig:PoiDisSheBiPlu2122Two} for \(h_{2}=5\) at different 
resolutions. 
As in the case of \(\msbi{F}_{\!\smash{\beta+}}^{(1)}\), the maximum error 
of 0.4\% in the largest component 
\(\lbrack\msbi{F}_{\smash{\beta+}}^{(2)}\rbrack_{122}\) 
of \(\msbi{F}_{\!\smash{\beta+}}^{(2)}\) 
is at the upper and lower boundaries of the region normal to \(\bm{i}_{2}\).
Since \(\msbi{F}_{\!\smash{\beta+}}^{(1)}\) determines 
\(\msbi{F}_{\!\smash{\beta+}}^{(2)}\), the corresponding boundary error 
is in the first two rows of points adjacent to the boundary. In contrast to 
the results for \(\lbrack\msbi{F}_{\smash{\beta+}}^{(1)}\rbrack_{12}\), 
there is no difference in the maximum error of 0.4\% between 
\(h_{2}=10\) and \(h_{2}=5\). 

\begin{figure}[H]
\centering
\includegraphics[width=0.4\textwidth]{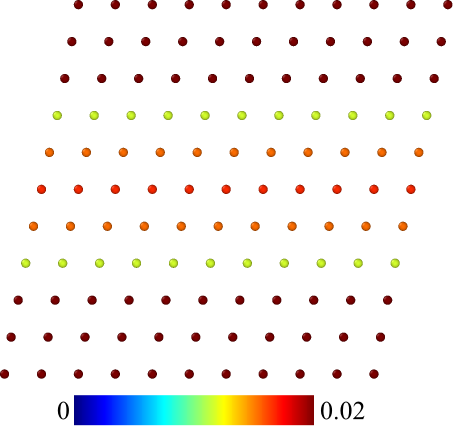}
\hspace{5mm}
\includegraphics[width=0.4\textwidth]{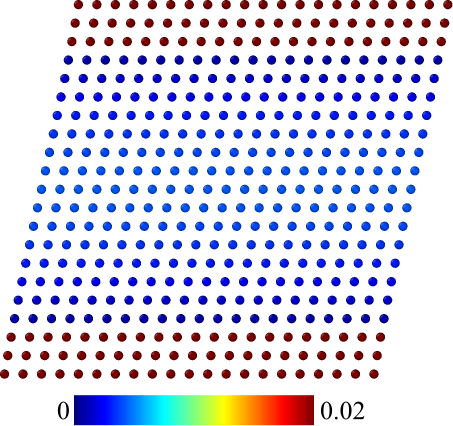}
\caption{Error 
\(
|h_{\smash{2}}^{2}
\lbrack\msbi{F}_{\smash{\beta+}}^{(3)}\rbrack_{1222}
-(\varsigma_{1}/h_{2})\,f^{\prime\prime\prime}
((\bm{i}_{2}\cdot\bm{r}_{\smash{\beta\mathrm{r}}}-c_{2})/h_{2})|
\) 
for \(h_{2}=10\) and two resolutions.\newline 
Left:~\(a_{0}=1\), \(n_{1,2}=11\). 
Right:~\(a_{0}=\frac{1}{2}\), \(n_{1,2}=21\). 
}
\label{fig:PoiDisSheBiPlu31222One}
\end{figure}

\begin{figure}[H]
\centering
\includegraphics[width=0.4\textwidth]{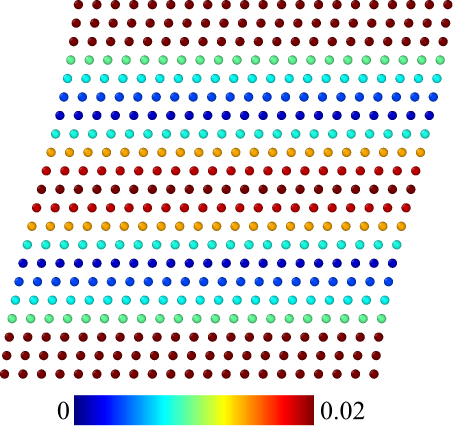}
\hspace{5mm}
\includegraphics[width=0.4\textwidth]{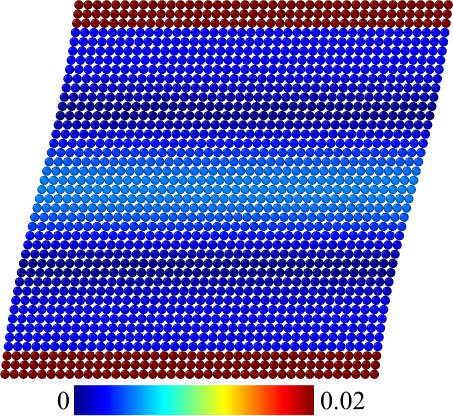}
\caption{Error \(|h_{\smash{2}}^{2}
\lbrack\msbi{F}_{\smash{\beta+}}^{(3)}\rbrack_{1222}
-(\varsigma_{1}/h_{2})
\,f^{\prime\prime\prime}
((\bm{i}_{2}\cdot\bm{r}_{\smash{\beta\mathrm{r}}}-c_{2})/h_{2})|\) 
for \(h_{2}=5\) and two resolutions.\newline 
Left:~\(a_{0}=\frac{1}{2}\), \(n_{1,2}=21\).
Right:~\(a_{0}=\frac{1}{4}\), \(n_{1,2}=41\). 
}
\label{fig:PoiDisSheBiPlu31222Two}
\end{figure}

Lastly, analogous results for the largest component 
\(\lbrack\msbi{F}_{\smash{\beta+}}^{(3)}\rbrack_{1222}\) 
of \(\msbi{F}_{\!\smash{\beta+}}^{(3)}\) 
are presented in Figure \ref{fig:PoiDisSheBiPlu31222One} for \(h_{2}=10\)
and in Figure \ref{fig:PoiDisSheBiPlu31222Two} for \(h_{2}=5\) at different 
resolutions. 
As evident, these trends are consistent with those just discussed 
for \(\msbi{F}_{\!\smash{\beta+}}^{(1)}\) and 
\(\msbi{F}_{\!\smash{\beta+}}^{(2)}\). As shown by this 
example, then, discrete local deformation measures of order \(2\) 
and greater characterize the non-affinity of discrete displacement data. 

\subsection{Displacement of atoms in a dislocated lattice} 
\label{sec:DisDisDL}

\begin{figure}[H]
\centering
\includegraphics[width=0.49\textwidth]{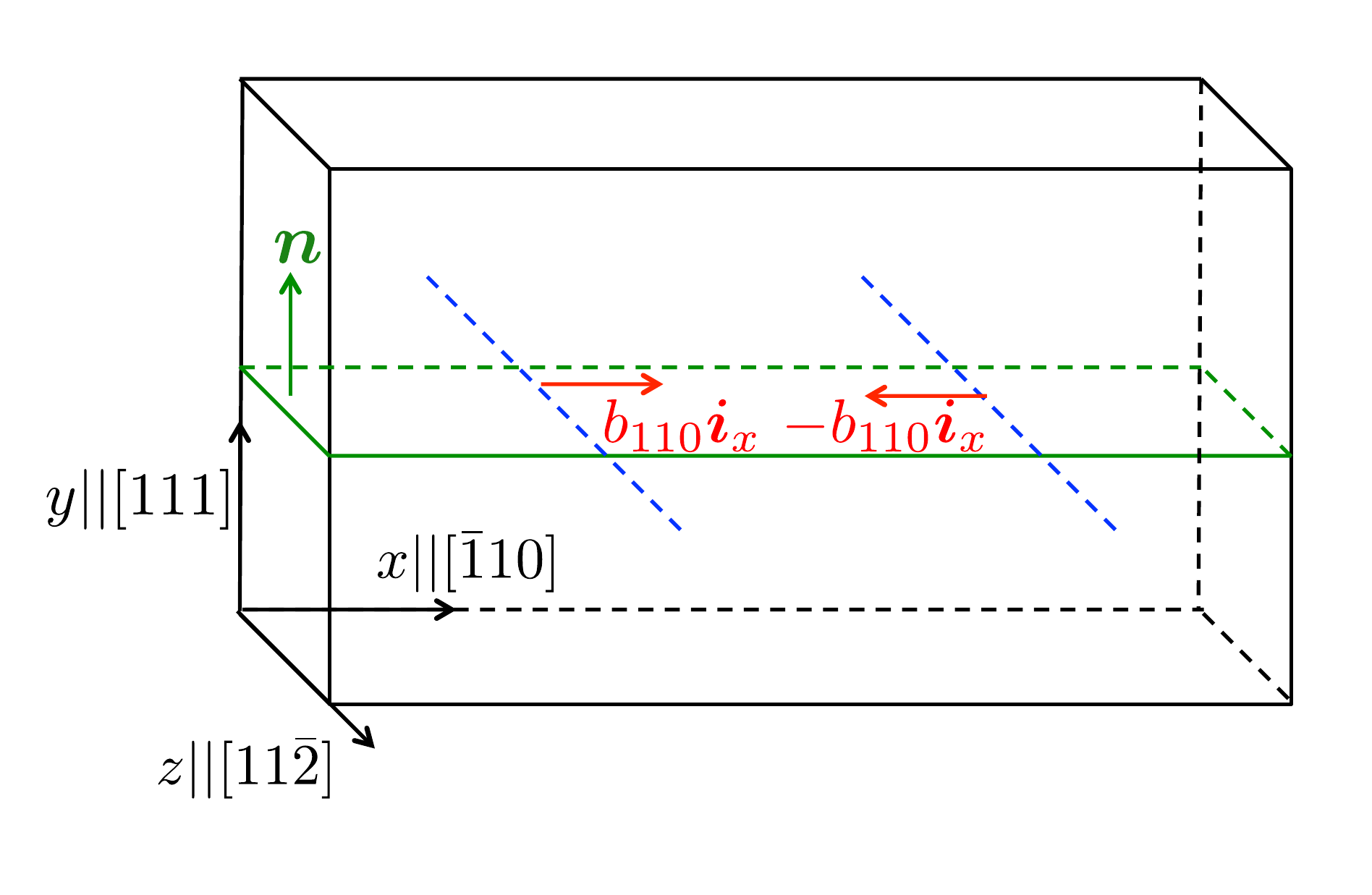}
%\hspace{5mm}
\includegraphics[width=0.45\textwidth]{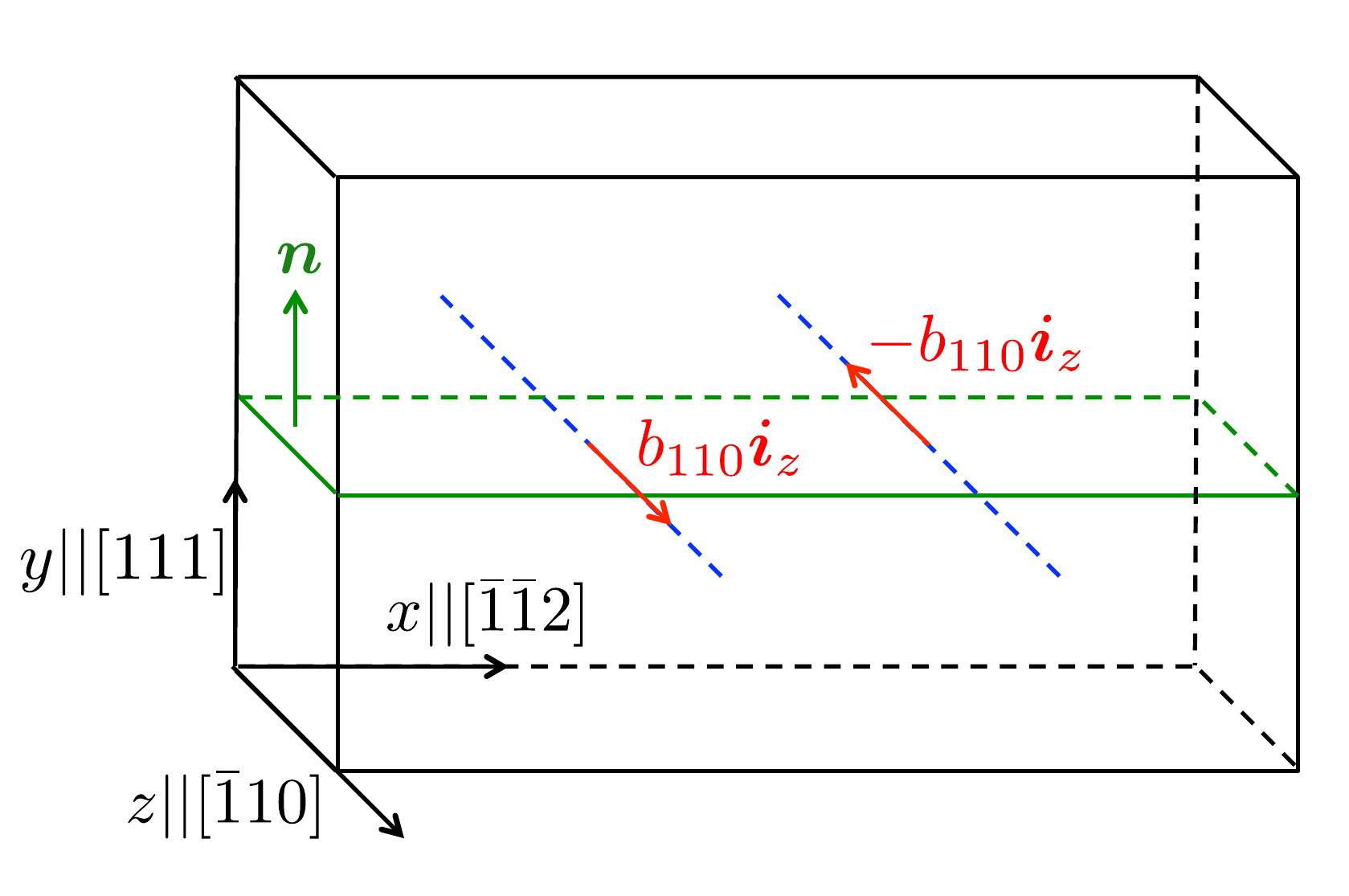}
\caption{Cells for MS simulation of atomic displacement due to a 
dissociated dislocation dipole. 
Initial configuration:~perfect dislocations at \((x,y)=(L_{x}/4,L_{y}/2)\) and 
\((3L_{x}/4,L_{y}/2)\) with Burgers vectors 
\(\pm\bm{b}=\pm a_{0}\lbrack\bar{1}10\rbrack/2\), 
\(b_{110}=|\bm{b}|=a_{0}/\sqrt{2}\). 
Left:~edge dipole with cell size 
\(
(L_{x},L_{y},L_{z})=(120\sqrt{6},180,9\sqrt{2}/2)\,d_{111}
\), 
\(d_{111}=a_{0}/\sqrt{3}\). 
Right:~screw dipole with cell size
\(
(L_{x},L_{y},L_{z})=(180\sqrt{2},180,3\sqrt{6})\,d_{111}
\). 
Both cells contain \(259200\) atoms. 
}
\label{fig:CellBoth}
\end{figure}

In this context, the set of points in question are \(n\) atoms with positions 
\(\bm{r}_{\!1},\ldots,\bm{r}_{\!n}\) in a crystallographic lattice 
containing a dislocation. Position results are obtained from molecular 
statics (MS) simulation of dislocation dipole relaxation. The simulation 
cells are shown in Figure \ref{fig:CellBoth}. 
Rather than Al and Cu as considered by \cite{Hartley2005a}, 
Au is employed here. 
Simulations are initialized via conjugate gradient relaxation and 
quadratic line search under constant (zero) stress and constant (0 K) 
temperature conditions in LAMMPS \cite[][]{Plimpton1995} via the 
box/relax command. Initially perfect dipoles are introduced by applying 
continuum displacements from linear elastic (i.e., Volterra) dislocation 
theory to core atoms \cite[e.g.,][Chapter 5]{Bul06}.Dissociation of these 
is simulated via 5000 steps of fast inertial relaxation 
\cite[using FIRE:][]{Bitzek2006} 
followed by 5000 steps of conjugate gradient relaxation, at zero stress. 

\begin{figure}[ht]
\centering
\includegraphics[width=0.45\textwidth]{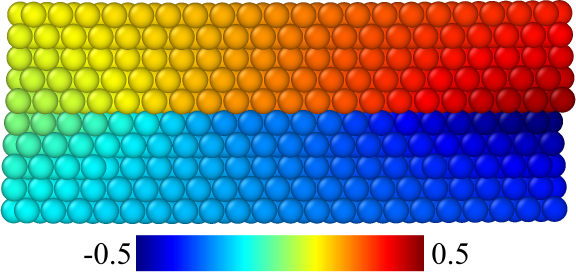}
%\hspace{5mm}
\includegraphics[width=0.45\textwidth]{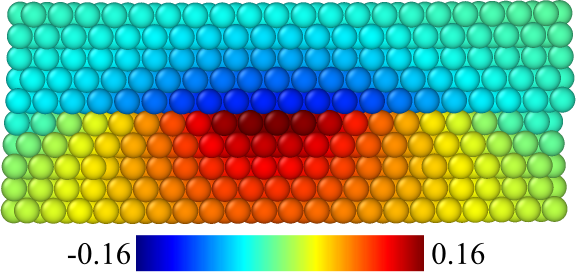}
\caption{Normalized atomic displacement results 
\(\bm{u}_{\alpha}/|\bm{b}|\cdot\bm{i}_{x}\) (left) and 
\(\bm{u}_{\alpha}/|\bm{b}|\cdot\bm{i}_{z}\) (right) in a region around 
the dissociated edge monopole at \((x,y)=(L_{x}/4,L_{y}/2)\) in 
Figure \ref{fig:CellBoth} (left) with Burgers vector 
\(\bm{b}=|\bm{b}|\,\bm{i}_{x}=b_{110}\,\bm{i}_{x}\). 
Regions above and below the glide plane are clearly visible. 
}
\label{fig:DisEdgDis}
\end{figure}

Displacement components for the dissociated left edge monopole in 
Figure \ref{fig:CellBoth} (left) are shown in Figure \ref{fig:DisEdgDis}.
Here and in what follows, all results are displayed at atoms in the \((x,y)\) 
plane perpendicular to the dislocation line, i.e., to \(\bm{i}_{z}\). 
Introducing a straight dislocation with Burgers vector \(\bm{b}\) into the bulk 
lattice on a glide plane results in a displacement 
\(
\bm{u}_{\alpha}(t)
:=\bm{r}_{\!\alpha}(t)-\bm{r}_{\!\alpha}(0)
\) 
of \(\alpha=1,\ldots,n\), with 
\(
-\frac{1}{2}
\leqslant
\bm{u}_{\alpha}/|\bm{b}|\cdot\bm{b}/|\bm{b}|
\leqslant
\frac{1}{2}
\). 
This is shown in Figure \ref{fig:DisEdgDis} (left). 
The displacement results in Figure \ref{fig:DisEdgDis} determine 
the position information employed to obtain all discrete local 
deformation results in the rest of this section. 

In contrast to the examples in Section \ref{sec:ExpValDL} for the case 
of a finite regular grid / simple cubic lattice, all points in an infinite 
periodic lattice are interior. For atoms with fcc neighborhoods, 
\(r_{\smash{\mathrm{N}s}}=\sqrt{s/2}\,a_{0}\) for \(s=1,2,3\), such 
that 
\(|S_{\!\smash{1\beta}}|=12\), 
\(|S_{\!\smash{2\beta}}|=18\), 
and \(|S_{\!\smash{3\beta}}|=42\). Consequently, 
\(|S_{\!\smash{d\beta}}|>3^{d}\) for \(d=1,2,3\). 
As indicated by the CNA analysis of the dislocated lattice 
(see Figure \ref{fig:LisNeiHM} below), atoms in the stacking fault (red) 
have hcp neighborhoods, and those in the partial dislocation cores (white) 
have triclinic (disordered) neighborhoods; all remaining atoms have fcc 
neighborhoods. Recall that the cut-off radius 
\(
r_{\smash{\mathrm{CNA}}}
=\frac{1}{2}(r_{\smash{\mathrm{N}1}}+r_{\smash{\mathrm{N}2}})
\) 
is employed in CNA to determine neighborhoods.

In what follows, \(\msbi{F}_{\!\smash{\beta-}}^{(d)}\) for \(d=1,2\) 
are determined for all atoms based on displacement results in Figure 
\ref{fig:DisEdgDis}. To this end, 
the algorithm of \cite{Hartley2005a} is employed. 
Reduction of \(n_{\smash{\beta-}}^{(1)}\) based on their algorithm 
is shown in Figure \ref{fig:LisNeiHM}. 
\begin{figure}[H]
\centering
\includegraphics[width=0.4\textwidth]{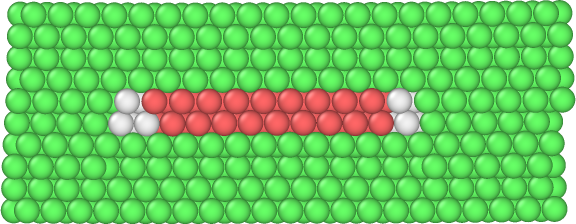}\\[2mm]
\includegraphics[width=0.4\textwidth]{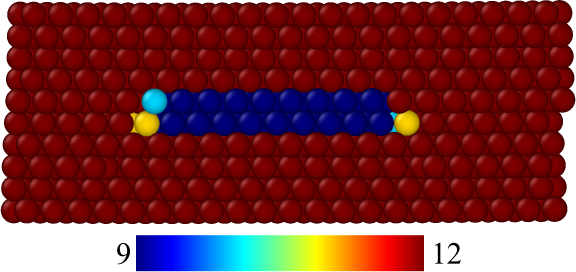}
\caption{Above:~common neighbor analysis (CNA) of atoms in the 
region around the dissociated edge monopole at \((x,y)=(L_{x}/4,L_{y}/2)\) 
in Figure \ref{fig:CellBoth} (left) showing atoms with lattice (fcc, green), 
stacking fault (hcp, red), and core (white, disordered) neighborhoods. 
Below:~reduced value of \(n_{\smash{\beta-}}^{(1)}\) based on the 
algorithm of \cite{Hartley2005a}. 
}
\label{fig:LisNeiHM}
\end{figure} 
Note the reduction of \(n_{\smash{\beta-}}^{(1)}\) below 
(fcc nearest neighbor value) 12 only for atoms in 
the stacking fault (red) or the partial dislocation cores (white). 

\subsubsection{Results for edge case}

\begin{figure}[H]
\centering
\includegraphics[width=0.45\textwidth]{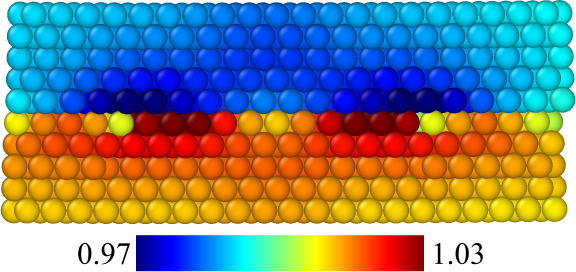}
%\hspace{5mm}
\includegraphics[width=0.45\textwidth]{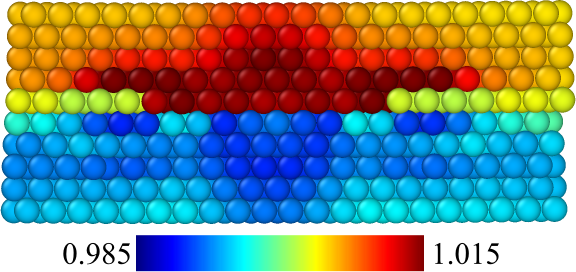}
\caption{Discrete local deformation of atomic neighborhoods 
in and around a dissociated edge dislocation core:~
\(\lbrack\msbi{F}_{\!\smash{\beta-}}^{(1)}\rbrack_{xx}\) 
(left) 
and 
\(\lbrack\msbi{F}_{\!\smash{\beta-}}^{(1)}\rbrack_{yy}\) 
(right). 
}
\label{fig:PosAtoEdgComNorF1}
\end{figure}

\begin{figure}[ht]
\centering
\includegraphics[width=0.45\textwidth]{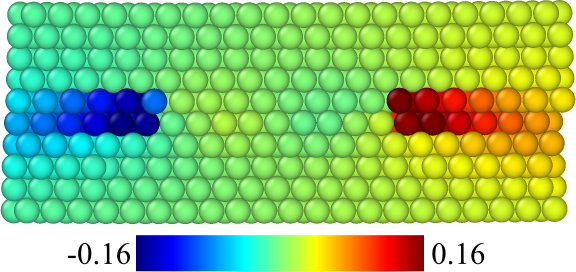}
%\hspace{5mm}
\includegraphics[width=0.45\textwidth]{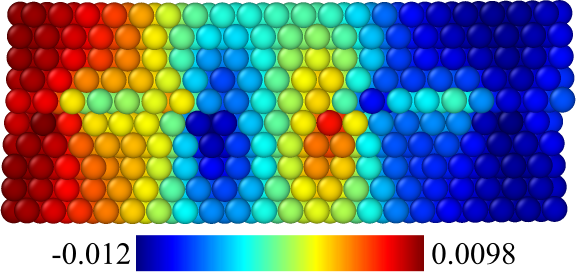}
\caption{Discrete local deformation of atomic neighborhoods 
in and around a dissociated edge dislocation core:~
\(\lbrack\msbi{F}_{\!\smash{\beta-}}^{(1)}\rbrack_{xy}\) 
(left) and 
\(\lbrack\msbi{F}_{\!\smash{\beta-}}^{(1)}\rbrack_{yx}\) 
(right). 
}
\label{fig:PosAtoEdgComSheF1}
\end{figure}

To begin, consider the components of \(\msbi{F}_{\!\smash{\beta-}}^{(1)}\) 
for the edge case. All results are displayed at atoms in the \((x,y)\) 
plane perpendicular to the dislocation line, i.e., to \(\bm{i}_{z}\), with 
the Burgers vector in the horizontal \(\bm{i}_{x}\) direction. 
Normal components of this are shown in Figure \ref{fig:PosAtoEdgComNorF1}, 
and shear components in Figure \ref{fig:PosAtoEdgComSheF1}. 
Comparison of these results with the CNA-based visualization in Figure 
\ref{fig:LisNeiHM} (above) shows that maximum normal and shear local 
deformation (distortion) is associated with the extended defect. Note also 
that \(\lbrack\msbi{F}_{\!\smash{\beta-}}^{(1)}\rbrack_{xy}\) in 
Figure \ref{fig:PosAtoEdgComSheF1} (left) is extremal for the core atoms. 

\begin{figure}[H]
\centering
\includegraphics[width=0.45\textwidth]{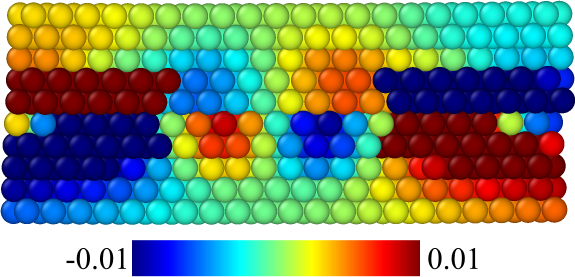}
%\hspace{5mm}
\includegraphics[width=0.45\textwidth]{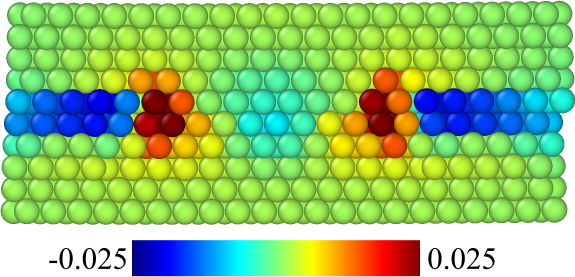}
\caption{Discrete local deformation of atomic neighborhoods 
in and around a dissociated edge dislocation core:~
\(d_{111}\lbrack\msbi{F}_{\!\smash{\beta-}}^{(2)}\rbrack_{xyy}\) 
(left) and 
\(
d_{111}
\lbrack
\mathrm{sym}_{2}\msbi{F}_{\!\smash{\beta-}}^{(2)}
\rbrack_{xxy}
\)
(right). 
Note \(d_{111}=a_{0}/\sqrt{3}=2.36\times 10^{-10}\) m for Au 
with \(a_{0}=4.08\times 10^{-10}\) m. 
}
\label{fig:PosAtoEdgF2}
\end{figure}

\begin{figure}[ht]
\centering
\includegraphics[width=0.45\textwidth]{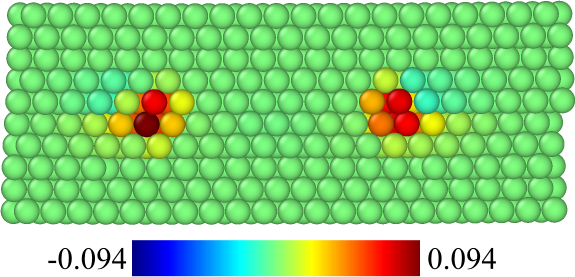}
%\hspace{5mm}
\includegraphics[width=0.45\textwidth]{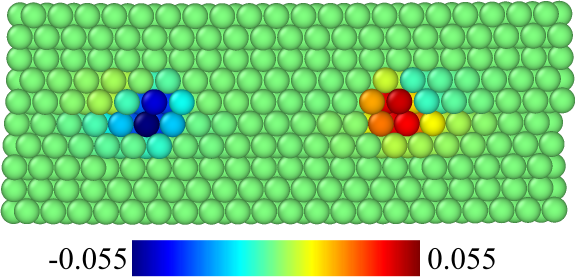}
\caption{Discrete local deformation of atomic neighborhoods 
in and around a dissociated edge dislocation core:~
\(
d_{111}\lbrack\msbi{G}_{\smash{\beta-}}^{(1)}\rbrack_{xz}
\) 
(left) 
and 
\(
d_{111}\lbrack\msbi{G}_{\smash{\beta-}}^{(1)}\rbrack_{zz}
\) 
(right). 
}
\label{fig:PosAtoEdgG1}
\end{figure}

Consider next the largest components of 
\(\msbi{F}_{\!\smash{\beta-}}^{(2)}\) and 
\(
\mathrm{sym}_{2}\msbi{F}_{\!\smash{\beta-}}^{(2)}
\) 
in Figure \ref{fig:PosAtoEdgF2}. 
In particular, 
\(
\lbrack\mathrm{sym}_{2}\msbi{F}_{\!\smash{\beta-}}^{(2)}\rbrack_{xxy}
=\tfrac{1}{2}
\,\lbrack\msbi{F}_{\!\smash{\beta-}}^{(2)}\rbrack_{xxy}
+\tfrac{1}{2}
\,\lbrack\msbi{F}_{\!\smash{\beta-}}^{(2)}\rbrack_{xyx}
\). 
Additional components of \(\msbi{F}_{\!\smash{\beta-}}^{(2)}\) determine the 
largest components 
\begin{equation}
\lbrack\msbi{G}_{\smash{\beta-}}^{(1)}\rbrack_{xz}
=\lbrack\msbi{F}_{\!\smash{\beta-}}^{(2)}\rbrack_{xyz}
-\lbrack\msbi{F}_{\!\smash{\beta-}}^{(2)}\rbrack_{xxy}
\,,\quad
\lbrack\msbi{G}_{\smash{\beta-}}^{(1)}\rbrack_{zz}
=\lbrack\msbi{F}_{\!\smash{\beta-}}^{(2)}\rbrack_{zyx}
-\lbrack\msbi{F}_{\!\smash{\beta-}}^{(2)}\rbrack_{zxy}
\label{equ:TenDisDisScrEdg}
\end{equation}
of its "axial vector" 
\(
\msbi{G}_{\smash{\beta-}}^{(1)}
:=\mathrm{axv}_{2}
\,2\,\mathrm{skw}_{2}\msbi{F}_{\!\smash{\beta-}}^{(2)}
\) 
via \eqref{equ:AxiSkwSwiDL} shown in Figure \ref{fig:PosAtoEdgG1}. 
Clearly, both 
\(\mathrm{sym}_{2}\msbi{F}_{\!\smash{\beta-}}^{(2)}\) and 
\(\msbi{G}_{\smash{\beta-}}^{(1)}\) are non-trivial for 
atoms in the neighborhood of a dissociated edge monopole in fcc Au.  

\subsubsection{Results for screw case} 

\begin{figure}[H]
\centering
\includegraphics[width=0.45\textwidth]{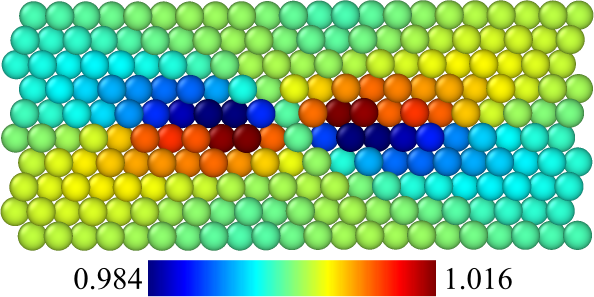}
%\hspace{5mm}
\includegraphics[width=0.45\textwidth]{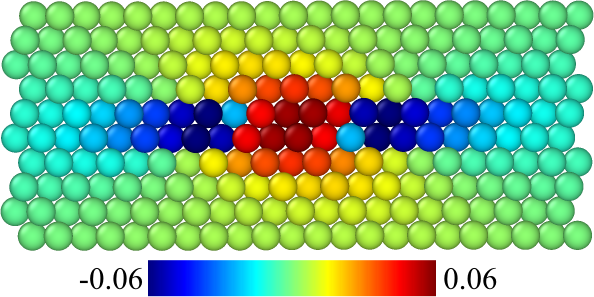}
\caption{Discrete local deformation of atomic neighborhoods 
in and around a dissociated screw dislocation core:~
\(\lbrack\msbi{F}_{\!\smash{\beta-}}^{(1)}\rbrack_{xx}\) 
(left) 
and 
\(\lbrack\msbi{F}_{\!\smash{\beta-}}^{(1)}\rbrack_{xy}\) 
(right).}
\label{fig:PosAtoScrComNorF1}
\end{figure}

\begin{figure}[H]
\centering
\includegraphics[width=0.45\textwidth]{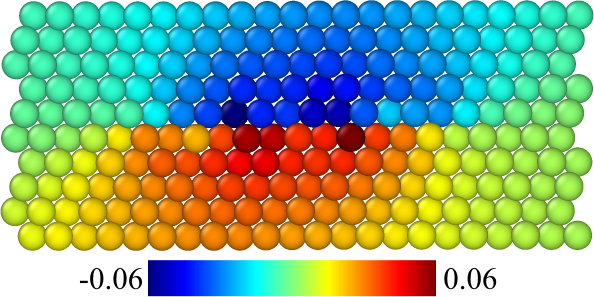}
%\hspace{5mm}
\includegraphics[width=0.45\textwidth]{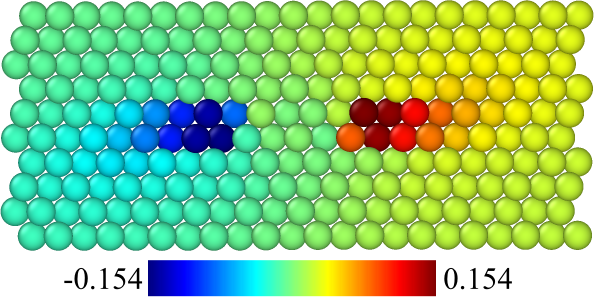}
\caption{Discrete local deformation of atomic neighborhoods in 
and around a dissociated screw dislocation core:~
\(\lbrack\msbi{F}_{\!\smash{\beta-}}^{(1)}\rbrack_{zx}\) 
(left) 
and 
\(\lbrack\msbi{F}_{\!\smash{\beta-}}^{(1)}\rbrack_{zy}\) 
(right).}
\label{fig:PosAtoScrComSheF1}
\end{figure}

Analogous results are obtained for discrete local deformation in atomic 
neighborhoods in a region surrounding the dissociated screw monopole 
at \((x,y)=(L_{x}/4,L_{y}/2)\) in Figure \ref{fig:CellBoth} (right). 
Again, all results are displayed at atoms in the \((x,y)\) 
plane perpendicular to the dislocation line, i.e., to \(\bm{i}_{z}\); now, 
however, the Burgers vector is oriented in the \(\bm{i}_{z}\) 
direction perpendicular to this plane. The largest components of 
\(\msbi{F}_{\!\smash{\beta-}}^{(1)}\) are shown in 
Figures \ref{fig:PosAtoScrComNorF1} and \ref{fig:PosAtoScrComSheF1}. 
Analogous to \(\lbrack\msbi{F}_{\!\smash{\beta-}}^{(1)}\rbrack_{xy}\) 
in the edge case in Figure \ref{fig:PosAtoEdgComSheF1} (left), the shear 
\(\lbrack\msbi{F}_{\!\smash{\beta-}}^{(1)}\rbrack_{zy}\) in the Burgers 
vector direction (\(\bm{i}_{z}\)) perpendicular to the glide plane shown in 
Figure \ref{fig:PosAtoScrComSheF1} (right) is the largest component 
in the screw case as well. Since the dominant partial Burgers vector 
component is edge-like in Figure \ref{fig:PosAtoEdgComSheF1}, 
and screw-like in Figure \ref{fig:PosAtoScrComSheF1}, note that 
\(\lbrack\msbi{F}_{\!\smash{\beta-}}^{(1)}\rbrack_{zy}\) is much 
more localized to the core atoms than 
\(\lbrack\msbi{F}_{\!\smash{\beta-}}^{(1)}\rbrack_{xy}\). 

\begin{figure}[H]
\centering
\includegraphics[width=0.49\textwidth]{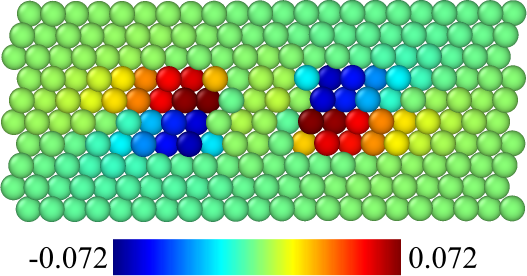}
\includegraphics[width=0.49\textwidth]{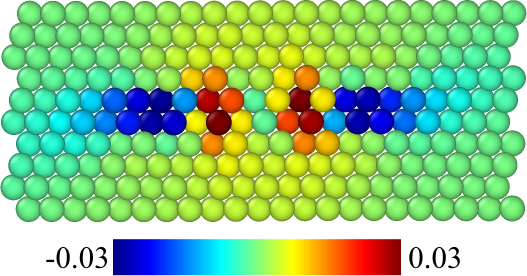}
\caption{Discrete local deformation of atomic neighborhoods 
in and around a dissociated screw dislocation core:~
\(d_{111}\lbrack\msbi{F}_{\!\smash{\beta-}}^{(2)}\rbrack_{zyy}\) 
(left) 
and 
\(d_{111}
\lbrack
\mathrm{sym}_{2}\msbi{F}_{\!\smash{\beta-}}^{(2)}
\rbrack_{zxy}\) 
(right).
}
\label{fig:PosAtoScrF2}
\end{figure}

\begin{figure}[H]
\centering
\includegraphics[width=0.49\textwidth]{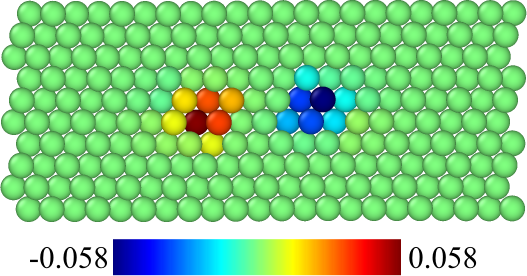}
\includegraphics[width=0.49\textwidth]{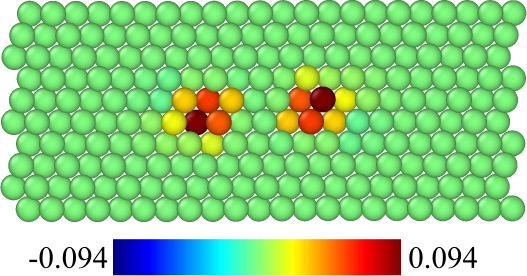}
\caption{Discrete local deformation 
of atomic neighborhoods in and around a dissociated screw dislocation core:~
\(d_{111}\lbrack\msbi{G}_{\smash{\beta-}}^{(1)}\rbrack_{xz}\) 
(left) 
and 
\(d_{111}\lbrack\msbi{G}_{\smash{\beta-}}^{(1)}\rbrack_{zz}\) 
(right). 
}
\label{fig:PosAtoScrG1}
\end{figure}

Likewise analogous to the edge case and results for 
\(\msbi{F}_{\!\smash{\beta-}}^{(2)}\) in Figure \ref{fig:PosAtoEdgF2} 
are those in Figure \ref{fig:PosAtoScrF2} for the screw case. 
Note that 
\(
\lbrack
\mathrm{sym}_{2}\msbi{F}_{\!\smash{\beta-}}^{(2)}
\rbrack_{zxy}
=\tfrac{1}{2}
\,\lbrack\msbi{F}_{\!\smash{\beta-}}^{(2)}\rbrack_{zxy}
+\tfrac{1}{2}
\,\lbrack\msbi{F}_{\!\smash{\beta-}}^{(2)}\rbrack_{zyx}
\). 
Again, since the dominant partial Burgers vector component is edge-like 
in Figure \ref{fig:PosAtoEdgF2}, and screw-like in Figure \ref{fig:PosAtoScrF2}, 
\(\lbrack\msbi{F}_{\!\smash{\beta-}}^{(2)}\rbrack_{zyy}\) is much 
more localized to the core atoms (i.e., partial dislocation lines) than 
\(\lbrack\msbi{F}_{\!\smash{\beta-}}^{(2)}\rbrack_{xyy}\). 
Lastly, Figure \ref{fig:PosAtoScrG1} displays the largest components 
of \(\msbi{G}_{\smash{\beta-}}^{(1)}\) for the screw case. 
These are the same components as those \eqref{equ:TenDisDisScrEdg} in 
the edge case. As expected, 
\(
\lbrack\msbi{G}_{\smash{\beta-}}^{(1)}\rbrack_{xz}
>\lbrack\msbi{G}_{\smash{\beta-}}^{(1)}\rbrack_{zz}
\) 
in the edge case in Figure \ref{fig:PosAtoEdgG1}, and this reverses in 
the screw case shown in Figure \ref{fig:PosAtoScrG1}. Indeed, 
\(
\lbrack\msbi{G}_{\smash{\beta-}}^{(1)}\rbrack_{xz}
\) 
is the edge component, and 
\( 
\lbrack\msbi{G}_{\smash{\beta-}}^{(1)}\rbrack_{zz}
\) 
the screw component, of \(\msbi{G}_{\smash{\beta-}}^{(1)}\) for these 
cases, as we now discuss in more detail. 

\section{Selected fields determined by discrete local deformation} 
\label{sec:DLDIntComPse}

\subsection{Basic considerations}

The elements of a discrete local deformation of order \(m\) 
\eqref{equ:DefLocDis} induce a number of different fields. 
For example, in the simplest case, we have 
\begin{equation}
\begin{array}{rcl}
\bm{\chi}_{\bm{c}}^{\smash{(2)}}(\bm{x})
&:=&
\bm{\chi}_{\bm{c}}^{\smash{(2)}}(\bm{c})
+\msbi{F}^{(1)}(\bm{x}-\bm{c})
+\tfrac{1}{2}
\,(\msbi{F}^{(2)}(\bm{x}-\bm{c}))\,(\bm{x}-\bm{c})
\,\\
\bm{F}_{\!\bm{c}}^{(2)}(\bm{x})
&:=&
\msbi{F}^{(1)}+\msbi{F}^{(2)}(\bm{x}-\bm{c})
\,,
\end{array}
\label{equ:FieDefLocGloSecOrd}
\end{equation} 
based on \((\msbi{F}^{(1)},\msbi{F}^{(2)})\) and "centered" at 
\(\bm{c}\in\mathbb{E}^{3}\). Given the identity 
\begin{equation}
(\cdots(\msbi{F}^{(d)}
\underbrace{\bm{a})\cdots)\,\bm{a}}\nolimits_{k\ \mathrm{times}}
=(\cdots((\mathrm{sym}_{k}\msbi{F}^{(d)})
\underbrace{\bm{a})\cdots)\,\bm{a}}\nolimits_{k\ \mathrm{times}}
\,,\ \ 
k\leqslant d
\,,
\label{equ:IdeSymGen}
\end{equation}
based on \eqref{equ:SkwSymGen}, one sees that 
\eqref{equ:FieDefLocGloSecOrd}${}_{1}$ 
depends in fact only on the completely symmetric part 
\(\mathrm{sym}_{2}\msbi{F}^{(2)}\) of \(\msbi{F}^{(2)}\), i.e., 
\begin{equation}
\bm{\chi}_{\bm{c}}^{\smash{(2)}}(\bm{x})
=\bm{\chi}_{\bm{c}}^{\smash{(2)}}(\bm{c})
+\msbi{F}^{(1)}(\bm{x}-\bm{c})
+\tfrac{1}{2}
\,((\mathrm{sym}_{2}\msbi{F}^{(2)})\,(\bm{x}-\bm{c}))\,(\bm{x}-\bm{c})
\,.
\label{equ:FieDefGloSecOrd}
\end{equation} 
In this case, the difference 
\begin{equation}
\begin{array}{rcl}
\bm{F}_{\!\bm{c}}^{(2)}(\bm{x})
-\nabla\bm{\chi}_{\bm{c}}^{\smash{(2)}}(\bm{x})
&=&
(\mathrm{skw}_{2}\msbi{F}^{(2)})\,(\bm{x}-\bm{c})
\end{array}
\label{equ:FieDefGloGraLocDifSecOrd}
\end{equation} 
is determined by the completely skew-symmetric part 
\(\mathrm{skw}_{2}\msbi{F}^{(2)}\) of \(\msbi{F}^{(2)}\). 
In addition, 
\begin{equation}
\begin{array}{rclcl}
\mathop{\mathrm{curl}}\nabla\bm{\chi}_{\bm{c}}^{\smash{(2)}}
&=&
\mathrm{axv}_{\mathrm{2}}
\,2\,\mathrm{skw}_{2}
\,\mathrm{sym}_{2}\msbi{F}^{(2)}
&=&
\bm{0}
\,,\\
\mathop{\mathrm{curl}}\bm{F}_{\!\bm{c}}^{(2)}
&=&
\mathrm{axv}_{\mathrm{2}}
\,2\,\mathrm{skw}_{2}\msbi{F}^{(2)}
&\neq&
\bm{0}
\,,
\end{array}
\label{equ:FieDefLocGloGraCurSecOrd}
\end{equation}
hold via \eqref{equ:SkwSymSwiDL}${}_{2}$, \eqref{equ:AxiSkwSwiDL} 
and \eqref{equ:CurTenOrdSecGenOne}. 

Examples of \eqref{equ:FieDefLocGloSecOrd}${}_{2}$ include those 
\begin{equation}
\bm{F}_{\!\smash{\beta\mathrm{r}}}(\bm{x}_{\mathrm{r}})
:=\msbi{F}_{\smash{\beta+}}^{(1)}
+\msbi{F}_{\smash{\beta+}}^{(2)}
(\bm{x}_{\mathrm{r}}-\bm{r}_{\!\smash{\beta\mathrm{r}}})
\,,\quad
\bm{F}_{\!\smash{\beta\mathrm{c}}}^{-1}(\bm{x}_{\mathrm{c}})
:=\msbi{F}_{\smash{\beta-}}^{(1)}
+\msbi{F}_{\smash{\beta-}}^{(2)}
(\bm{x}_{\mathrm{c}}-\bm{r}_{\!\smash{\beta\mathrm{c}}})
\,,
\label{equ:FieDefLocDisSecOrd}
\end{equation}
determined by 
\((\msbi{F}_{\smash{\beta\pm}}^{(1)},\msbi{F}_{\smash{\beta\pm}}^{(2)})\) 
from Section \ref{sec:CorDisAppDL}. In this context, we have 
\begin{equation}
\begin{array}{rclcl}
\mathrm{curl}_{\mathrm{r}}
\bm{F}_{\!\beta\mathrm{r}}
&=&
\mathrm{axv}_{\mathrm{2}\,}
2\,\mathrm{skw}_{2}
\msbi{F}_{\!\smash{\beta+}}^{(2)}
&=:&
\msbi{G}_{\smash{\beta+}}^{(1)}
\,,\\
\mathrm{curl}_{\mathrm{c}}
\bm{F}_{\!\smash{\beta\mathrm{c}}}^{-1}
&=&
\mathrm{axv}_{\mathrm{2}\,}
2\,\mathrm{skw}_{2}
\msbi{F}_{\!\smash{\beta-}}^{(2)}
&=:&
\msbi{G}_{\smash{\beta-}}^{(1)}
\,.
\end{array}
\label{equ:FieDefLocDisSecOrdCur}
\end{equation} 
In particular, note the formal analogy of 
\(
\mathrm{curl}_{\mathrm{c}}\bm{F}_{\!\smash{\beta\mathrm{c}}}^{-1}
\) 
to 
\(
\mathrm{curl}_{\mathrm{c}}\bm{F}_{\!\smash{\mathrm{L}}}^{-1}
\) 
in the context of the multiplicative decompostion 
\(\nabla\bm{\chi}=\bm{F}_{\!\mathrm{L}}\bm{F}_{\!\mathrm{R}}\) 
of the continuum deformation gradient into lattice 
\(\bm{F}_{\!\mathrm{L}}\) and residual (e.g., dislocation) 
\(\bm{F}_{\!\mathrm{R}}\) contributions. 
As well known in continuum dislocation theory \cite[e.g.,][]{Cer01}, 
\(
\mathrm{curl}_{\mathrm{c}}\bm{F}_{\!\smash{\mathrm{L}}}^{-1}
\) 
is a finite deformation generalization of the Nye dislocation tensor 
\cite[][]{Nye53}. 

Consider next the fields 
\begin{equation}
\begin{array}{rcl}
\bm{\chi}_{\bm{c}}^{\smash{(3)}}(\bm{x})
&:=&
\bm{\chi}_{\bm{c}}^{\smash{(3)}}(\bm{c})
+\msbi{F}^{(1)}(\bm{x}-\bm{c})
+\tfrac{1}{2}
\,((\mathrm{sym}_{2}\msbi{F}^{(2)})\,(\bm{x}-\bm{c}))\,(\bm{x}-\bm{c})
\\
&+&
\tfrac{1}{3}
\,(((\mathrm{sym}_{3}\msbi{F}^{(3)})\,(\bm{x}-\bm{c}))
\,(\bm{x}-\bm{c}))\,(\bm{x}-\bm{c})
\,,\\
\bm{F}_{\!\bm{c}}^{(3)}(\bm{x})
&:=&
\msbi{F}^{(1)}
+\msbi{F}^{(2)}(\bm{x}-\bm{c})
+((\mathrm{sym}_{2}\msbi{F}^{(3)})\,(\bm{x}-\bm{c}))\,(\bm{x}-\bm{c})
\,,
\end{array}
\label{equ:FieDefLocGloThrOrd}
\end{equation} 
centered at \(\bm{c}\) and determined by 
\((\msbi{F}^{(1)},\msbi{F}^{(2)},\msbi{F}^{(3)})\) 
via \eqref{equ:IdeSymGen}. In this case, 
\begin{equation}
\begin{array}{rclcl}
\mathop{\mathrm{curl}}\nabla\bm{\chi}_{\bm{c}}^{\smash{(3)}}
&=&
\mathrm{axv}_{\mathrm{2}}
\,2\,\mathrm{skw}_{2}
\,\lbrack
\mathrm{sym}_{2}\msbi{F}^{(2)}
+2\,(\mathrm{sym}_{3}\msbi{F}^{(3)})\,(\bm{x}-\bm{c})
\rbrack
&=&
\bm{0}
\,,\\
\mathop{\mathrm{curl}}\bm{F}_{\!\bm{c}}^{(3)}
&=&
\mathrm{axv}_{\mathrm{2}}
\,2\,\mathrm{skw}_{2}
\,\lbrack
\msbi{F}^{(2)}
+2\,(\mathrm{sym}_{2}\msbi{F}^{(3)})\,(\bm{x}-\bm{c})
\rbrack
&\neq&
\bm{0}
\,,
\end{array}
\label{equ:FieDefLocGloGraCurThrOrd}
\end{equation}
hold analogous to \eqref{equ:FieDefLocGloGraCurSecOrd}. Further, 
\begin{equation}
\bm{F}_{\!\bm{c}}^{(3)}(\bm{x})
-\nabla\bm{\chi}_{\bm{c}}^{\smash{(3)}}(\bm{x})
=(\mathrm{skw}_{2}\msbi{F}^{(2)})\,(\bm{x}-\bm{c})
+((\mathrm{sym}_{2,3}\msbi{F}^{(3)})\,(\bm{x}-\bm{c}))\,(\bm{x}-\bm{c})
\label{equ:FieDefGloGraLocDifThrOrd}
\end{equation}
is determined not only by \(\mathrm{skw}_{2}\msbi{F}^{(2)}\) as in 
\eqref{equ:FieDefGloGraLocDifSecOrd} above, but also by the difference 
\begin{equation}
\mathrm{sym}_{d-1,d}\msbi{F}^{(d)}
:=\mathrm{sym}_{d-1}\msbi{F}^{(d)}-\mathrm{sym}_{d}\msbi{F}^{(d)}
\label{equ:FieDefGloGraLocDifSym}
\end{equation}
for \(d=3\). This is also true for the general deformation fields 
\begin{equation}
\begin{array}{rcl}
\bm{\chi}_{\bm{c}}^{\smash{(m)}}(\bm{x})
&=&
\bm{\chi}_{\!\bm{c}}^{\smash{(m)}}(\bm{c})
+\msbi{F}^{(1)}(\bm{x}-\bm{c})
+\tfrac{1}{2}
\,((\mathrm{sym}_{2}\msbi{F}^{(2)})\,(\bm{x}-\bm{c}))\,(\bm{x}-\bm{c})
+\cdots
\\
&+&
\tfrac{1}{m}
\,(\cdots((\mathrm{sym}_{m}\msbi{F}^{(m)})
\underbrace{(\bm{x}-\bm{c}))\cdots)\,(\bm{x}-\bm{c})}
\nolimits_{m\,\mathrm{times}}
\,,\\
\bm{F}_{\!\bm{c}}^{(m)}(\bm{x})
&:=&
\msbi{F}^{(1)}
+\msbi{F}^{(2)}(\bm{x}-\bm{c})
+((\mathrm{sym}_{2}\msbi{F}^{(3)})\,(\bm{x}-\bm{c}))\,(\bm{x}-\bm{c})
+\cdots
\\
&+&
(\cdots((\mathrm{sym}_{m-1}\msbi{F}^{(m)})
\underbrace{(\bm{x}-\bm{c}))\cdots)\,(\bm{x}-\bm{c})}
\nolimits_{(m-1)\,\mathrm{times}}
\,,
\end{array}
\label{equ:FieDefLocGloGenOrd}
\end{equation} 
induced by all elements of \eqref{equ:DefLocDis}, i.e., 
\begin{equation}
\begin{array}{rcl}
\bm{F}_{\!\bm{c}}^{(m)}
-\nabla\bm{\chi}_{\!\bm{c}}^{\smash{(m)}}(\bm{x})
&=&
(\mathrm{skw}_{2}\msbi{F}^{(2)})\,(\bm{x}-\bm{c})
+((\mathrm{sym}_{2,3}\msbi{F}^{(3)})\,(\bm{x}-\bm{c}))\,(\bm{x}-\bm{c})
+\cdots
\\
&+&
(\cdots((\mathrm{sym}_{m-1,m}\msbi{F}^{(m)})
\underbrace{(\bm{x}-\bm{c}))\cdots)\,(\bm{x}-\bm{c})}
\nolimits_{(m-1)\,\mathrm{times}}
\,.
\end{array}
\label{equ:FieDefGloGraLocDifGenOrd}
\end{equation} 
Like in the case of \(\bm{\chi}_{\bm{c}}^{\smash{(2,3)}}\) discussed above, 
\(\mathop{\mathrm{curl}}\nabla\bm{\chi}_{\bm{c}}^{\smash{(m)}}=\bm{0}\), 
again via \eqref{equ:IdeSymGen}. 
Besides the difference between the two measures, 
the right-hand sides of \eqref{equ:FieDefGloGraLocDifSecOrd}, 
\eqref{equ:FieDefGloGraLocDifThrOrd} and 
\eqref{equ:FieDefGloGraLocDifGenOrd} represent the information 
contained in \eqref{equ:DefLocDis} which is lost when one works with 
\(\bm{\chi}_{\bm{c}}^{\smash{(m)}}\) alone. 

\subsection{Generalization of local subset DIC}
\label{sec:SLDICGen}

In the current notation, local subset DIC is based on the model form 
\begin{equation}
\textstyle
\bm{\chi}_{\mathrm{DIC}}(\bm{x}_{\mathrm{r}})
:=\sum_{i}\kappa_{i}(\bm{x}_{\mathrm{r}})
\,\bm{\chi}_{\smash{\bm{x}_{i\mathrm{r}}}}^{(1)}(\bm{x}_{\mathrm{r}})
\,,
\label{equ:DefGloSLDIC}
\end{equation} 
for the continuum deformation field of a region of interest 
\(R=\bigcup_{i}R_{i}\), with 
\begin{equation}
\bm{\chi}_{\smash{\bm{x}_{i\mathrm{r}}}}^{(1)}
(\bm{x}_{\mathrm{r}})
:=\bm{\chi}_{\smash{\bm{x}_{i\mathrm{r}}}}^{(1)}
(\bm{x}_{i\mathrm{r}})
+\msbi{F}_{\smash{i+}}^{(1)}(\bm{x}_{\mathrm{r}}-\bm{x}_{i\mathrm{r}})
\,,\quad
\bm{x}_{i\mathrm{c}}
=\bm{\chi}_{\smash{\bm{x}_{i\mathrm{r}}}}^{\smash{(1)}}
(\bm{x}_{i\mathrm{r}})
\,.
\label{equ:DefGloOneOrdROI}
\end{equation} 
Assume now that 
\(
(\msbi{F}_{\!\smash{i+}}^{(1)},\ldots,\msbi{F}_{\!\smash{i+}}^{(m)})
\) 
has been determined for each \(R_{i}\subset R\). In this case, 
\eqref{equ:DefGloSLDIC} generalizes directly to 
\begin{equation}
\textstyle
\bm{\chi}_{\mathrm{DIC}}(\bm{x}_{\mathrm{r}})
:=\sum_{i}\kappa_{i}(\bm{x}_{\mathrm{r}})
\,\bm{\chi}_{\!\smash{\bm{x}_{i\mathrm{r}}}}^{(m)}(\bm{x}_{\mathrm{r}})
\label{equ:DefGloSLDICGen}
\end{equation} 
via \eqref{equ:FieDefLocGloGenOrd}${}_{1}$. In addition, 
local subset DIC can be generalized to the determination of
\begin{equation}
\textstyle
\bm{F}_{\mathrm{DIC}}(\bm{x}_{\mathrm{r}})
:=\sum_{i}\kappa_{i}(\bm{x}_{\mathrm{r}})
\,\bm{F}_{\!\smash{\bm{x}_{i\mathrm{r}}}}^{(m)}(\bm{x}_{\mathrm{r}})
\label{equ:DefLocSLDICGen}
\end{equation} 
as well via \eqref{equ:FieDefLocGloGenOrd}${}_{2}$. Besides being 
of interest in its own right, extending or generalizing local subset DIC 
in this fashion accounts for the additional physical information contained in 
\(
(\msbi{F}_{\!\smash{i+}}^{(1)},\ldots,\msbi{F}_{\!\smash{i+}}^{(m)})
\) 
on generally non-affine and / or incompatible local deformation in real 
materials. 

\section{Relation of current treatment to selected previous work}
\label{sec:WorPreSel} 

Besides to \cite{Hartley2005a} and \cite{Shimizu2007}, 
the current treatment is most closely related to that of \cite{Zimmerman2009}. 
In particular, these latter authors work with 
the generalization 
\begin{equation}
\bm{s}_{\alpha\beta}(t)
\approx
\msbi{F}_{\!\smash{\beta+}}^{(1)}(t)\,\bm{s}_{\alpha\beta}(0)
+\tfrac{1}{2}
\,(\msbi{S}_{\smash{\beta+}}^{(2)}(t)\,\bm{s}_{\alpha\beta}(0))
\,\bm{s}_{\alpha\beta}(0)
\label{equ:DefLocZim}
\end{equation} 
of the first-order case 
\eqref{equ:VecDefSepDefLocOrdOne}${}_{1}$ (i.e., their Equation (18)), 
where 
\(
\msbi{S}_{\smash{\beta+}}^{(2)}(t)
\in
\mathrm{Sym}_{2,2}(\mathbb{V}^{3},\mathbb{V}^{3})
\). 
In addition, note that their Equation (40) corresponds to 
\eqref{equ:VecDefSepDefLocTwoOrd}${}_{1}$ for 
\(\msbi{F}_{\!\smash{\beta+}}^{(2)}\). These authors 
do not consider \(\msbi{F}_{\!\smash{\beta-}}^{(1)}\) 
from \eqref{equ:VecDefSepDefLocOrdOne}${}_{2}$ and 
\(\msbi{F}_{\!\smash{\beta-}}^{(2)}\) from 
\eqref{equ:VecDefSepDefLocTwoOrd}${}_{2}$. As mentioned above, 
the first of these was treated by \cite{Hartley2005a}. 

Although somewhat different in purpose, the more recent work of \cite{Zhang2015} 
is conceptually related to the current treatment as well. 
These authors employ a weighted least-squares fit of 
\begin{equation}
\begin{array}{rclcl}
\bm{r}_{\!\alpha}(t)
&\approx&
\bm{\chi}_{\mathrm{Z}t}(\bm{r}_{\!\alpha}(0))
&=&
\bm{\chi}_{\mathrm{Z}t}(\bm{x}_{\mathrm{r}}+\bm{s}_{\alpha}(0))
\\
&&&\approx&
\bm{\chi}_{\mathrm{Z}t}(\bm{x}_{\mathrm{r}})
+\nabla_{\!\smash{\mathrm{r}}}^{(1)}
\bm{\chi}_{\mathrm{Z}t}(\bm{x}_{\mathrm{r}})
\,\bm{s}_{\alpha}(0)
+\tfrac{1}{2!}
\,(\nabla_{\!\smash{\mathrm{r}}}^{(2)}
\bm{\chi}_{\mathrm{Z}t}(\bm{x}_{\mathrm{r}})
\,\bm{s}_{\alpha}(0))
\bm{s}_{\alpha}(0)
\end{array}
\label{equ:FieDefGloZha}
\end{equation}
(based on their Equations (12)-(20) and (33)-(43)) to atomic position results 
\(\bm{r}_{\!\alpha}\), \(\alpha=1,\ldots,n\), from molecular dynamics (MD) 
at the position \(\bm{x}_{\mathrm{r}}\in B_{\mathrm{r}}\) in a 
reference configuration \(B_{\mathrm{r}}\subset\mathbb{E}^{3}\). 
In the process, best-fit values of
\begin{equation}
\bm{x}_{\mathrm{c}}
=\bm{\chi}_{\mathrm{Z}t}
(\bm{x}_{\mathrm{r}})
\in
\mathbb{V}^{3}
\,,\ \
\nabla_{\!\smash{\mathrm{r}}}^{(1)}
\bm{\chi}_{\mathrm{Z}t}(\bm{x}_{\mathrm{r}})
\in
\mathrm{Lin}_{1}(\mathbb{V}^{3},\mathbb{V}^{3})
\,,\ \ 
\nabla_{\!\smash{\mathrm{r}}}^{(2)}
\bm{\chi}_{\mathrm{Z}t}(\bm{x}_{\mathrm{r}})
\in
\mathrm{Sym}_{2,2}(\mathbb{V}^{3},\mathbb{V}^{3})
\,,
\label{equ:FitDefZha}
\end{equation}
are determined for all atoms (i.e., in the simulation cell) at 
\(\bm{x}_{\mathrm{r}}\in B_{\mathrm{r}}\). 

As implied by the dislocation core example in Section \ref{sec:DisDisDL}, 
even for the displacement of atoms in the neighborhood of a single atom
located at \(\bm{r}_{\beta}(0)=\bm{x}_{\mathrm{r}}\), 
both \eqref{equ:DefLocZim} and \eqref{equ:FieDefGloZha} may be 
qualitatively too special. 
Indeed, in contrast to the analogous discrete local deformation 
\((\msbi{F}_{\!\smash{\beta+}}^{(1)}(t),
\msbi{F}_{\!\smash{\beta+}}^{(2)}(t))\), neither 
\(\msbi{S}_{\smash{\beta+}}^{(2)}(t)\) 
nor 
\(
\nabla_{\!\smash{\mathrm{r}}}^{(2)}
\bm{\chi}_{\mathrm{Z}t}(\bm{x}_{\mathrm{r}})
\) 
captures any incompatible contribution from these displacements 
to continuum (local) deformation. Moreover, it is not clear that it 
makes physical sense in general to assume that there exists a 
\textit{single} continuum deformation field \(\bm{\chi}_{\mathrm{Z}t}\) 
which represents the displacements of \textit{all} atoms at any 
\(\bm{x}_{\mathrm{r}}\in B_{\mathrm{r}}\). Indeed, any 
deformation field like \(\bm{\chi}_{\mathrm{Z}t}\) 
is in fact just one element of an \textit{equivalence class} 
\(
\lbrack\bm{\chi}\rbrack_{\smash{\bm{x}_{\mathrm{r}}}}^{(2)}
\) 
of such fields at any \(\bm{x}_{\mathrm{r}}\in B_{\mathrm{r}}\); 
for the case of order \(m\), 
\(
\lbrack\bm{\chi}\rbrack_{\smash{\bm{x}_{\mathrm{r}}}}^{(m)}
\) 
is defined by\footnote{In differential geometry, 
this represents a so-called \(m\)-jet \cite[e.g.,][Chapter 6]{Kolar1993}; 
in a continuum mechanical setting, see also for example \cite{Mor75} or 
\cite{Sve09}.}  
\begin{equation}
\nabla_{\!\smash{\mathrm{r}}}^{(1)}
\bm{\chi}_{a}(\bm{x}_{\mathrm{r}})
=\nabla_{\!\smash{\mathrm{r}}}^{(1)}
\bm{\chi}_{b}(\bm{x}_{\mathrm{r}})
\,,
\ldots,
\nabla_{\!\smash{\mathrm{r}}}^{(m)}
\bm{\chi}_{a}(\bm{x}_{\mathrm{r}})
=\nabla_{\!\smash{\mathrm{r}}}^{(m)}
\bm{\chi}_{b}(\bm{x}_{\mathrm{r}})
\,,
\label{equ:ClaEquRefDefGlo}
\end{equation}
for all 
\(
\bm{\chi}_{a},\bm{\chi}_{b}
\in
\lbrack
\bm{\chi}
\rbrack_{\smash{\bm{x}_{\mathrm{r}}}}^{(m)}
\). 
Likewise, 
\begin{equation}
\nabla_{\!\smash{\mathrm{c}}}^{(1)}
\bm{\chi}_{a}^{\smash{-1}}(\bm{x}_{\mathrm{c}})
=\nabla_{\!\smash{\mathrm{c}}}^{(1)}
\bm{\chi}_{b}^{\smash{-1}}(\bm{x}_{\mathrm{c}})
\,,
\ldots,
\nabla_{\!\smash{\mathrm{c}}}^{(m)}
\bm{\chi}_{a}^{\smash{-1}}(\bm{x}_{\mathrm{c}})
=\nabla_{\!\smash{\mathrm{c}}}^{(m)}
\bm{\chi}_{b}^{\smash{-1}}(\bm{x}_{\mathrm{c}})
\,,
\label{equ:ClaEquCurDefGlo}
\end{equation}
for all 
\(
\bm{\chi}_{a}^{\smash{-1}},\bm{\chi}_{b}^{\smash{-1}}
\in
\lbrack
\bm{\chi}^{\smash{-1}}
\rbrack_{\smash{\bm{x}_{\mathrm{c}}}}^{(m)}
\) 
defines the equivalence class 
\(
\lbrack
\bm{\chi}^{\smash{-1}}
\rbrack_{\smash{\bm{x}_{\mathrm{c}}}}^{(m)}
\) 
at any position \(\bm{x}_{\mathrm{c}}\in B_{\mathrm{c}}\) in the 
current configuration \(B_{\mathrm{c}}\subset\mathbb{E}^{3}\). 
Since 
\(
\nabla_{\!\smash{\mathrm{r}}}^{(d)}
\bm{\chi}(\bm{x}_{\mathrm{r}})
\in 
\mathrm{Sym}_{d,d}(\mathbb{V}^{3},\mathbb{V}^{3})
\), 
note that 
the completely symmetric part 
\(
(\msbi{F}_{+}^{\smash{(1)}},
\mathrm{sym}_{2}\msbi{F}_{+}^{\smash{(2)}},
\ldots,
\mathrm{sym}_{m}\msbi{F}_{+}^{\smash{(m)}})
\) 
of the discrete local deformation  
\(
(\msbi{F}_{+}^{\smash{(1)}},
\msbi{F}_{+}^{\smash{(2)}},
\ldots,
\msbi{F}_{+}^{\smash{(m)}})
\) 
can represent 
\(
\lbrack
\bm{\chi}
\rbrack_{\smash{\bm{x}_{\mathrm{r}}}}^{(m)}
\). 
Similarly, 
\(
\lbrack
\bm{\chi}^{-1}
\rbrack_{\smash{\bm{x}_{\mathrm{c}}}}^{(m)}
\) 
can be represented by the completely symmetric part 
\(
(\msbi{F}_{-}^{\smash{(1)}},
\mathrm{sym}_{2}\msbi{F}_{-}^{\smash{(2)}},
\ldots,
\mathrm{sym}_{m}\msbi{F}_{-}^{\smash{(m)}})
\) 
of 
\(
(\msbi{F}_{-}^{\smash{(1)}},
\msbi{F}_{-}^{\smash{(2)}},
\ldots,
\msbi{F}_{-}^{\smash{(m)}})
\).  

\section{Summary and discussion} 
\label{sec:DisSumDL} 

In the current work, the concept of discrete local deformation 
of order \(m\) has been developed to characterize discrete 
displacement data in a pseudo-continuum kinematic fashion. 
Central to the current approach are the generalizations 
\eqref{equ:VecDefSepDefLocTwoOrd} and \eqref{equ:VecDefSepDefLocOrdI} 
of \eqref{equ:VecDefSepDefLocOrdOne}. Together with 
\eqref{equ:VecDefSepDefLocOrdOne}, these result in a hierarchical 
determination of discrete local deformation from discrete position 
information in the sense that order \(m\) depends on the results 
of all lower orders \(m-1,\ldots,1\). The least-squares-based 
over-determination of discrete local deformation measures gives these 
the character of spatially-averaged quantities in a (finite) neighborhood 
of each point. The current approach generalizes those of 
\cite{Hartley2005a}, \cite{Shimizu2007} and 
\cite{Zimmerman2009} for the determination of order \(m=1\) 
discrete local deformation from atomic displacement information to 
(i) arbitrary discrete displacement information and 
(ii) discrete local deformation measures of order \(m>2\). 
More specifically, \cite{Shimizu2007} worked explicitly with 
\eqref{equ:LagEulOrdOne}${}_{1}$ based on 
\eqref{equ:VecDefSepDefLocOrdOne}${}_{1}$, and 
\cite{Hartley2005a} with \eqref{equ:LagEulOrdOne}${}_{2}$ based on 
\eqref{equ:VecDefSepDefLocOrdOne}${}_{2}$. 
In addition, \cite{Zimmerman2009} discussed relations equivalent to 
\eqref{equ:LagEulTwoOrd}${}_{1}$ based on 
\eqref{equ:VecDefSepDefLocTwoOrd}${}_{1}$. 

%Consistent with continuum Volterra dislocation theory (e.g., Hirth and Lothe, 1982), 
%the edge results in Figures 14 and 15 can be interpreted as more plane-strain-like, 
%while the screw results in Figures 19 and 20 are more anti-plane-shear-like.

As implied by the results in Section \ref{sec:ExpValDL}, 
for a region of fixed spatial size (i.e., specimen size, simulation cell size), 
the determination of \(\msbi{F}_{\!\smash{\beta\pm}}^{(d)}\) for 
\(d=1,\ldots,m\) is influenced by the point spacing / number \(n\) of points. 
In the experimental context, this spacing is limited by the resolution 
of the measurement method (e.g., local subset DIC). 
In the lattice / atomistic context (e.g., Section \ref{sec:DisDisDL}), 
this spacing is atomic and so (physically) fixed. In both cases, 
the size of \(n_{\smash{\beta}}^{(d)}<n\) is also relevant. 
Since this is not known \textit{a-priori} (especially in the empirical, 
experimental context), a convergence study is indicated, formally 
analogous to that in the numerical solution of continuum boundary-value 
problems. 

In the approach developed in Section \ref{sec:DefLocDet3D}, 
the discrete local deformation measures 
\(
\msbi{F}_{\!\smash{\beta\pm}}^{(1)},
\ldots,
\msbi{F}_{\!\smash{\beta\pm}}^{(m)}
\) 
are determined from \(\bm{r}_{\!1},\ldots,\bm{r}_{\!n}\) 
in a completely "decoupled" fashion, representing the simplest 
approach. A more accurate, coupled determination, however, is also 
possible. For example, consider the "first-order" generalizations 
\begin{equation}
\begin{array}{rcl}
\bm{s}_{\alpha\beta\pm}
&\approx&
\msbi{F}_{\!\smash{\beta\pm}}^{(1)}\,\bm{s}_{\alpha\beta\pm}
+\tfrac{1}{2}\,((\mathrm{sym}_{2}\msbi{F}_{\smash{\beta\pm}}^{(2)})
\,\bm{s}_{\alpha\beta\pm})
\,\bm{s}_{\alpha\beta\pm}
\,,\\
\msbi{F}_{\smash{\alpha\pm}}^{(1)}
&\approx&
\msbi{F}_{\smash{\beta\pm}}^{(1)}
+\msbi{F}_{\smash{\beta\pm}}^{(2)}\,\bm{s}_{\alpha\beta\pm}
\,,
\end{array}
\end{equation}
of \eqref{equ:VecDefSepDefLocOrdOne} and \eqref{equ:VecDefSepDefLocTwoOrd} 
via \eqref{equ:DefLocDisFirOrdDif} 
with \(\bm{s}_{\alpha\beta-}:=\bm{s}_{\alpha\beta}(t)\) 
and \(\bm{s}_{\alpha\beta+}:=\bm{s}_{\alpha\beta}(0)\) 
for simultaneous determination of 
\(
\msbi{F}_{\!\smash{\beta\pm}}^{(1)}
\)
and
\(
\msbi{F}_{\!\smash{\beta\pm}}^{(2)}
\) 
for  
\(
\beta=1,\ldots,n
\).
The Euler-Lagrange relations of the corresponding generalization 
\begin{equation}
\begin{array}{l}
f(\msbi{F}_{\!\smash{1\pm}}^{(1)},
\ldots,
\msbi{F}_{\!\smash{n\pm}}^{(1)},
\msbi{F}_{\!\smash{1\pm}}^{(2)},
\ldots,
\msbi{F}_{\!\smash{n\pm}}^{(2)})
\\
\quad=\ 
\frac{1}{2}
\sum_{\beta=1}^{n}\sum_{\alpha=1_{\beta\pm}}^{n_{\beta\pm}}
|\bm{s}_{\alpha\beta\pm}
-\msbi{F}_{\!\smash{\beta\pm}}^{(1)}\bm{s}_{\alpha\beta\mp}
-\tfrac{1}{2}\,(\msbi{F}_{\smash{\beta\pm}}^{(2)}\bm{s}_{\alpha\beta\mp})
\,\bm{s}_{\alpha\beta\mp}|^{2}
\\
\quad+\ \ 
\frac{1}{2}
\sum_{\beta=1}^{n}\sum_{\alpha=1_{\beta\pm}}^{n_{\beta\pm}}
|\msbi{F}_{\smash{\alpha\pm}}^{(1)}
-\msbi{F}_{\smash{\beta\pm}}^{(1)}
-\msbi{F}_{\smash{\beta\pm}}^{(2)}\bm{s}_{\alpha\beta\mp}|^{2}
\,,
\end{array}
\end{equation}
of the least-squares-based objective function 
with \(n_{\beta\pm}:=n_{\smash{\beta\pm}}^{(1,2)}\) yield a couple system 
for \(\msbi{F}_{\!\smash{1\pm}}^{(1)},
\ldots,
\msbi{F}_{\!\smash{n\pm}}^{(1)},
\msbi{F}_{\!\smash{1\pm}}^{(2)},
\ldots,
\msbi{F}_{\!\smash{n\pm}}^{(2)}
\) 
which can be solved numerically. Starting values for the corresponding 
iterative solution are available from the decoupled determination of these 
developed in Section \ref{sec:DefLocDet3D}.  

The concept of discrete local deformation employed here as based on 
can be developed further in a number of directions. 
For example, note that \eqref{equ:DefLocDis} can be 
expressed in the "reduced" form\footnote{This is based on the "pull-back" 
\(
\msbi{F}^{(1)\ast}\msbi{F}^{(d)}
=\msbi{F}^{(1)\,-1}\msbi{F}^{(d)}
\) 
of \(\msbi{F}^{(d)}\) by \(\msbi{F}^{(1)}\). One could also work with the 
analogous "push-forward" \(\msbi{F}_{\smash{\ast}}^{(1)}\msbi{F}^{(d)}\); 
for example, 
\(
((\bm{F}_{\smash{\ast}}^{(1)}\msbi{F}^{(2)})\bm{a})\bm{b}
=(\msbi{F}^{(2)}\msbi{F}^{(1)\,-1}\bm{a})
\msbi{F}^{(1)\,-1}\bm{b}
\). 
} 
\begin{equation}
(\msbi{F}^{(1)},
\msbi{\Gamma}^{(2)},
\ldots,
\msbi{\Gamma}^{(m)})
\,,\quad
\msbi{\Gamma}^{(d)}
:=\msbi{F}^{(1)\,-1}\msbi{F}^{(d)}
\,.
\label{equ:DefLocRed}
\end{equation}
In the context of Section \ref{sec:DefLocDet3D}, for example, note that 
\(
\msbi{F}_{\!\smash{\beta\pm}}^{(1)}
\)
and 
\(
\msbi{F}_{\!\smash{\beta\pm}}^{(d)} 
\) 
determine 
\(
\msbi{\Gamma}_{\!\smash{\beta\pm}}^{(d)}
\) 
for \(d=2,\ldots,m\) via \eqref{equ:DefLocRed}. 
Referring again to the discrete position configuration 
\(\bm{r}_{\!1}(t),\ldots,\bm{r}_{\!n}(t)\) as current, and 
to \(\bm{r}_{\!1}(0),\ldots,\bm{r}_{\!n}(0)\) as referential, 
note that the discrete local deformation measures 
\(
\msbi{F}_{\!\smash{\beta\pm}}^{(d)} 
\) 
are mixed current-referential, whereas 
\(
\msbi{\Gamma}_{\!\smash{\beta+}}^{(d)}
\) 
is purely referential, and 
\(
\msbi{\Gamma}_{\!\smash{\beta-}}^{(d)}
\) 
purely current, in character. This is also the case for the corresponding fields. 
Local deformation fields induced by \eqref{equ:DefLocRed} include those 
\begin{equation}
\begin{array}{rcl}
\bm{D}_{\!\bm{c}}^{(2)}(\bm{x})
&:=&
\msbi{I}^{(1)}+\msbi{\Gamma}^{(2)}(\bm{x}-\bm{c})
\,,\\
\bm{D}_{\!\bm{c}}^{(d)}(\bm{x})
&:=&
\msbi{I}^{(1)}
+\msbi{\Gamma}^{(2)}(\bm{x}-\bm{c})
+((\mathrm{sym}_{2}\msbi{\Gamma}^{(3)})
\,(\bm{x}-\bm{c}))\,(\bm{x}-\bm{c})
+\cdots
\\
&+&
(\cdots((\mathrm{sym}_{d-1}\msbi{\Gamma}^{(d)})
\underbrace{(\bm{x}-\bm{c}))\cdots)\,(\bm{x}-\bm{c})}
\nolimits_{(d-1)\,\mathrm{times}}
\,,
\end{array}
\end{equation} 
for \(3\leqslant d\leqslant m\). Analogous to \eqref{equ:FieDefLocDisSecOrd}, 
for example, 
\begin{equation}
\bm{D}_{\beta\mathrm{r}}(\bm{x}_{\mathrm{r}})
:=\bm{I}+\msbi{\Gamma}_{\!\smash{\beta+}\,}^{(2)}
(\bm{x}_{\mathrm{r}}-\bm{r}_{\!\smash{\beta\mathrm{r}}})
\,,\quad
\bm{D}_{\smash{\beta\mathrm{c}}}^{(2)}(\bm{x}_{\mathrm{c}})
:=\bm{I}+\msbi{\Gamma}_{\!\smash{\beta-}\,}^{(2)}
(\bm{x}_{\mathrm{c}}-\bm{r}_{\!\smash{\beta\mathrm{c}}})
\,,
\end{equation}
are purely referential, and purely current, respectively, in character, 
in contrast to the mixed current-referential measures 
\(\bm{F}_{\!\beta\mathrm{r}}\) and 
\(\bm{F}_{\!\smash{\beta\mathrm{c}}}^{-1}\). Likewise, whereas 
\(
\mathrm{curl}_{\mathrm{r}}\bm{F}_{\!\beta\mathrm{r}}
=\msbi{F}_{\!\smash{\beta+}}^{(1)}
\,\mathrm{curl}_{\mathrm{r}}
\bm{D}_{\beta\mathrm{r}}
\) 
and 
\(
\mathrm{curl}_{\mathrm{c}}
\bm{F}_{\!\smash{\beta\mathrm{c}}}^{-1}
=\msbi{F}_{\!\smash{\beta-}}^{(1)}
\,\mathrm{curl}_{\mathrm{c}}
\bm{D}_{\smash{\beta\mathrm{c}}}
\)  
are mixed, 
\(
\mathrm{curl}_{\mathrm{r}}\bm{D}_{\!\beta\mathrm{r}}
\) 
is purely referential, and 
\(
\mathrm{curl}_{\mathrm{c}}
\bm{D}_{\!\smash{\beta\mathrm{c}}}
\) 
is purely current, in character. 

Lastly, from the point of view of differential geometry 
\cite[e.g.,][]{Abr88}, note that the elements 
\(
(\msbi{\Gamma}^{(2)},
\ldots,
\msbi{\Gamma}^{(m)})
\) 
of \eqref{equ:DefLocRed} induce the connection fields
\begin{equation}
\begin{array}{rcl}
\msbi{\Gamma}_{\!\bm{c}}^{(3)}(\bm{x})
&:=&
\msbi{\Gamma}^{(2)}
+\msbi{\Gamma}^{(3)}(\bm{x}-\bm{c})
\,,\\
\msbi{\Gamma}_{\!\bm{c}}^{(d)}(\bm{x})
&:=&
\msbi{\Gamma}^{(2)}
+\msbi{\Gamma}^{(3)}(\bm{x}-\bm{c})
+((\mathrm{sym}_{2}
\msbi{\Gamma}^{(4)})\,(\bm{x}-\bm{c}))\,(\bm{x}-\bm{c})
+\cdots
\\
&+&
(\cdots((\mathrm{sym}_{d-2}\msbi{\Gamma}^{(d)})
\underbrace{(\bm{x}-\bm{c}))\cdots)\,(\bm{x}-\bm{c})}
\nolimits_{(d-2)\,\mathrm{times}}
\,,
\end{array}
\label{equ:DefLocFieRed}
\end{equation}
with \(4\leqslant d\leqslant m\). These are characterized by their torsion 
\(
2\,\mathrm{skw}_{2\,}\msbi{\Gamma}_{\!\bm{c}}^{(d)}
\) 
and curvature 
\(
2\,\mathrm{skw}_{2\,}
(\nabla\msbi{\Gamma}_{\!\bm{c}}^{(d)}
+\msbi{\Gamma}_{\!\bm{c}}^{(d)}
\triangle
\msbi{\Gamma}_{\!\bm{c}}^{(d)})
\), 
where 
\(
((\msbi{A}^{(d)}\triangle\msbi{B}^{(d)})\bm{a})\bm{b}
:=(\msbi{A}^{(d)}\bm{a})(\msbi{B}^{(d)}\bm{b})
\). 
In particular, \(\msbi{\Gamma}^{(2)}\) can be 
interpreted as a constant (Koszul) connection with torsion 
\(
2\,\mathrm{skw}_{2\,}\msbi{\Gamma}^{(2)}
=\msbi{F}^{(1)\,-1}\mathop{\mathrm{axt}}\msbi{G}^{(1)}
\) 
and curvature
\(
2\,\mathrm{skw}_{2\,}
(\msbi{\Gamma}_{\!\bm{c}}^{(2)}
\triangle
\msbi{\Gamma}_{\!\bm{c}}^{(2)})
\). 
These and other such measures offer a more general, comprehensive 
characterization of dislocations and other defects (e.g., disclinations, 
interfaces), and more generally material microstructure, 
than that limited to the Nye tensor. As in the case of this latter, 
these can be compared with corresponding theoretical measures 
in the context of of defect theory, micro- and nanomechanics 
\cite[e.g.,][]{Teo82,Mur87,Li08}. 
These and other aspects of the current approach represent 
work in progress to be reported on in the future. 

\smallskip

\textsl{Acknowledgements.} 
Financial support by the German Science Foundation (DFG) 
in the Collaborative Research Center SFB 761 
is gratefully acknowledged. 

\bibliographystyle{elsarticle-harv}

\begin{appendix} 

\renewcommand{\thesection}{\Alph{section}}
\setcounter{section}{0}
\renewcommand{\theequation}{\thesection.\arabic{equation}}
\setcounter{equation}{0}
\renewcommand{\thefigure}{\thesection.\arabic{figure}}
\setcounter{figure}{0}

\section{Curl of a second-order tensor field}
\label{app:CurFieTenOrdSec}

As well-known in continuum mechanics \cite[e.g.,][Chapter 2]{Mal69}, 
there is no single convention for the divergence and curl of second- or 
higher-order three-dimensional Euclidean tensor fields. Consequently, 
quantities like the (Nye) dislocation tensor depend on the definition 
or convention chosen. 
Since this is also an issue in the work of \cite{Hartley2005a} as well 
as in the interpretation of discrete local deformation in terms of 
pseudo-continuum kinematics, a brief discussion of this is topic 
provided here. 

Let \(\bm{u}\) be a differentiable (Euclidean) vector field, and 
\(\bm{T}\) a differentiable second-order tensor field. 
Following for example \citet[][Chapter 1, Equation (90)]{Chadwick1999}, 
the curl of \(\bm{u}\) can be defined as the vector field 
\(\mathop{\mathrm{curl}}\bm{u}\) satisfying\footnote{By convention, 
all operators such as \(\nabla\), \(\mathop{\mathrm{div}}\) and 
\(\mathop{\mathrm{curl}}\) apply to everything on their right.}
\begin{equation} 
\bm{a}\cdot\mathop{\mathrm{curl}}\bm{u}
:=\mathop{\mathrm{div}}\bm{u}\times\bm{a}
\label{equ:FieVecCurChaDef}
\end{equation}
for all (constant) \(\bm{a}\). Given this, consider the definitions 
\begin{equation}
(\mathrm{curl}_{1}\bm{T})^{\mathrm{T}}\bm{c}
:=\mathop{\mathrm{curl}}\bm{T}^{\mathrm{T}}\!\bm{c}
=:(\mathrm{curl}_{2}\bm{T})\bm{c}
\,,\ \ 
(\mathrm{curl}_{3}\bm{T})\bm{c}
:=\mathop{\mathrm{curl}}\bm{T}\bm{c}
=:(\mathrm{curl}_{4}\bm{T})^{\mathrm{T}}\bm{c}
\,,
\label{equ:CurTenOrdSecAll}
\end{equation} 
of \(\mathop{\mathrm{curl}}\bm{T}\). In particular, 
\(
\mathrm{curl}_{2}\bm{T}
\) 
is common in the literature on micromechanics and dislocation field theory 
\cite[e.g.,][]{Kos79,Mur87,Cer01}. Here, we work with 
\begin{equation}
(\mathop{\mathrm{curl}}\bm{T})^{\mathrm{T}}\bm{c}
:=\mathop{\mathrm{curl}}\bm{T}^{\mathrm{T}}\!\bm{c}
\quad\Longrightarrow\quad
\mathop{\mathrm{curl}}\bm{T}
\equiv
\mathrm{curl}_{1}\bm{T}
\,,
\label{equ:CurTenOrdSec}
\end{equation} 
following for example \cite{Teo82} and \cite{Sve02}. In terms of this convention, 
note that 
\begin{equation}
\mathop{\mathrm{curl}}\bm{T}
=2\,\mathrm{axv}_{2}\mathrm{skw}_{2}\nabla\bm{T}
\label{equ:CurTenOrdSecGenOne} 
\end{equation} 
holds via \eqref{equ:SkwSymSwiDL}${}_{2}$ and 
\eqref{equ:AxiSkwSwiDL}. Indeed, given  
\(
\bm{u}\times(\bm{a}\times\bm{b})
=(\bm{b}\cdot\bm{u})\bm{a}-(\bm{a}\cdot\bm{u})\bm{b}
\), 
\eqref{equ:FieVecCurChaDef} implies 
\begin{equation}
\mathop{\mathrm{curl}}\bm{u}
\cdot
\bm{a}\times\bm{b}
=(\nabla_{\!\bm{a}}\bm{u})\cdot\bm{b}
-(\nabla_{\!\bm{b}}\bm{u})\cdot\bm{a}
=\bm{b}\cdot(2\mathop{\mathrm{skw}}\nabla\bm{u})\bm{a}
\label{equ:CurTenOrdFirSve}
\end{equation} 
in terms of \(\nabla_{\!\bm{a}}\bm{u}:=(\nabla\bm{u})\bm{a}\). 
Applying \eqref{equ:CurTenOrdFirSve} to 
\(\bm{u}=\bm{T}^{\mathrm{T}}\bm{c}\), 
\begin{equation}
\begin{array}{l}
(\mathop{\mathrm{curl}}\bm{T}^{\mathrm{T}}\bm{c})
\cdot
\bm{a}\times\bm{b}
=\nabla_{\!\bm{a}}(\bm{T}^{\mathrm{T}}\bm{c})\cdot\bm{b}
-\nabla_{\!\bm{b}}(\bm{T}^{\mathrm{T}}\bm{c})\cdot\bm{a}
\\
\quad=\ 
\bm{c}\cdot(\nabla_{\!\bm{a}}\bm{T})\bm{b}
-\bm{c}\cdot(\nabla_{\!\bm{b}}\bm{T})\bm{a}
=\bm{c}
\cdot
((2\,\mathrm{skw}_{2}\nabla\bm{T})\bm{a})\bm{b}
\,.
\end{array}
\label{equ:CurTenOrdFirSecSve}
\end{equation} 
On the basis of the definition \eqref{equ:CurTenOrdSec}, one then obtains 
\begin{equation}
(\mathop{\mathrm{curl}}\bm{T})\,(\bm{a}\times\bm{b})
=(\nabla_{\!\bm{a}}\bm{T})\bm{b}-(\nabla_{\!\bm{b}}\bm{T})\bm{a}
=((2\,\mathrm{skw}_{2}\nabla\bm{T})\bm{a})\bm{b}
\,,
\label{equ:CurTenOrdSecGen} 
\end{equation} 
and so \eqref{equ:CurTenOrdSecGenOne} via the fact that 
\(
(\mathop{\mathrm{curl}}\bm{T})\,(\bm{a}\times\bm{b})
=(\mathop{\mathrm{curl}}\bm{T})\,(\mathop{\mathrm{axt}}\bm{a})\bm{b}
\). 

Consider lastly Cartesian component relations. Let  
\(
\bm{i}_{i}\times\bm{i}_{\!j}
=\epsilon_{i\!jk}\bm{i}_{k}
\) 
as usual. For \eqref{equ:CurTenOrdSecAll}${}_{1}$, we have 
\begin{equation}
\lbrack\mathrm{curl}_{1}\bm{T}\rbrack_{i\!j}
=\bm{i}_{i}
\cdot
(\mathrm{curl}_{1}\bm{T})\,\bm{i}_{\!j}
=\bm{i}_{\!j}
\cdot
\mathop{\mathrm{curl}}\bm{T}^{\mathrm{T}}\bm{i}_{i}
=\mathop{\mathrm{div}}T_{ik}\epsilon_{k\!jl}\,\bm{i}_{l}
=T_{ik,l}\epsilon_{k\!jl}
\label{equ:ComCarCurTenOrdSec} 
\end{equation}
via \eqref{equ:FieVecCurChaDef} for the Cartesian component form 
of \(\mathop{\mathrm{curl}}\bm{T}\). Likewise, 
\begin{equation}
\lbrack\mathrm{curl}_{2}\bm{T}\rbrack_{i\!j}
=\epsilon_{kil}T_{\!jk,l}
\,,\quad
\lbrack\mathrm{curl}_{3}\bm{T}\rbrack_{i\!j}
=T_{ki,l}\epsilon_{k\!jl}
\,,\quad
\lbrack\mathrm{curl}_{4}\bm{T}\rbrack_{i\!j}
=\epsilon_{kil}T_{k\!j,l}
\,.
\label{equ:ComCarCurTenOrdSecTwo} 
\end{equation} 
for \eqref{equ:CurTenOrdSecAll}${}_{2-4}$. 
Comparison of these with the definitions 
\begin{equation}
\begin{array}{rclcl}
\nabla\times\bm{T}
&:=&
T_{\!k\!j,l}
\,\bm{i}_{l}\times\bm{i}_{k}\otimes\bm{i}_{\!j}
&=&
\epsilon_{lki}T_{\!k\!j,l}\,\bm{i}_{i}\otimes\bm{i}_{\!j}
\,,\\
\bm{T}\times\nabla
&:=&
T_{\!ik,l}
\,\bm{i}_{i}\otimes\bm{i}_{k}\times\bm{i}_{l}
&=&
T_{\!ik,l}\epsilon_{kl\!j}\,\bm{i}_{i}\otimes\bm{i}_{\!j}
\,, 
\end{array}
\label{equ:FielTenSecOrdMalDef}
\end{equation}
from \citet[][Equations (2.5.36) and (2.5.38), respectively]{Mal69} 
for example implies the correspondences 
\begin{equation}
\begin{array}{rclclcl}
\lbrack\nabla\times\bm{T}\rbrack_{i\!j}
&=&
\epsilon_{lki}T_{k\!j,l}
&=&
\epsilon_{kil}T_{k\!j,l}
&=&
\lbrack\mathrm{curl}_{4}\bm{T}\rbrack_{i\!j}
\,,\\
\lbrack\bm{T}\times\nabla\rbrack_{i\!j}
&=&
T_{ik,l}\epsilon_{kl\!j}
&=&
-T_{ik,l}\epsilon_{k\!jl}
&=&
-\lbrack\mathrm{curl}_{1}\bm{T}\rbrack_{i\!j}
\,,
\label{equ:CurMalCor}
\end{array}
\end{equation}
between the respective component forms. 

\section{Determination of the Nye tensor in \cite{Hartley2005a}}
\label{app:TenNyeHM}

\citet[][\S 3.3]{Hartley2005a} employ the notation 
\(
\mathbf{P}^{(\beta)}
\equiv
\mathbf{S}_{\smash{\beta}}^{(1)\mathrm{T}}(0)
\),
\(
\mathbf{Q}^{(\beta)}
\equiv
\mathbf{S}_{\smash{\beta}}^{(1)\mathrm{T}}(t)
\) 
and 
\(
\mathbf{G}^{(\beta)}
\equiv
\mathbf{F}_{\smash{\beta-}}^{(1)\mathrm{T}}(t)
\). 
They work with the array 
\(
\mathbf{g}_{\smash{kl}}^{(\beta)}
:=(g_{\smash{kl}}^{(1\beta)},
\ldots,
g_{\smash{kl}}^{(n\beta)})
\), 
where
\(
g_{\smash{kl}}^{(\alpha\beta)}
:=G_{\smash{kl}}^{(\alpha)}-G_{\smash{kl}}^{(\beta)}
\) 
(i.e., their Equation (19)). Then the correspondence
\(
g_{\smash{kl}}^{(\alpha\beta)}
\equiv
\lbrack\msbi{H}_{\smash{\alpha\beta-}}^{(1)}(t)\rbrack_{lk}
\) 
holds via \eqref{equ:DefLocDisFirOrdDif}. 
Introducing next \(\mathbf{a}_{\smash{kl}}^{(\beta)}\) via 
\(
\mathbf{g}_{\smash{kl}}^{(\beta)}
=\mathbf{Q}_{\smash{\beta}}
\mathbf{a}_{\smash{kl}}^{(\beta)}
\) 
(their Equation (20)), inversion of this latter relation yields 
\(
\mathbf{a}_{\smash{kl}}^{(\beta)}
=(\mathbf{Q}_{\smash{\beta}}^{\mathrm{T}}\mathbf{Q}_{\beta})^{-1}
\mathbf{Q}_{\smash{\beta}}^{\mathrm{T}}\mathbf{g}_{\smash{kl}}^{(\beta)}
\) 
(corresponding to their Equation (21)). Rather than proceeding in a purely 
discrete fashion as done in the current work, \cite{Hartley2005a} (tacitly) 
introduce the field \(\skew3\hat{\bm{G}}^{(\beta)}(\bm{x}_{\mathrm{c}})\) 
with 
\(
\bm{G}^{(\beta)}
=\skew3\hat{\bm{G}}^{(\beta)}(\bm{r}_{\!\beta}(t))
\), 
assume 
\(
\nabla^{\mathrm{c}}\skew3\hat{\bm{G}}^{(\beta)}
\approx
\bm{i}_{k}\otimes\bm{i}_{l}\otimes\bm{a}_{\smash{kl}}^{(\beta)}
\), 
and (somehow) use the definition 
\(
\hat{\bm{\alpha}}
:=-\nabla\times\hat{\bm{G}}
\) 
(i.e., their Equation (11), leaving off the superscript \(\beta\)) of the Nye tensor. 
The component form of this is given by 
\(
\hat{\alpha}_{i\!j}
=-\lbrack\nabla\times\hat{\bm{G}}\rbrack_{i\!j}
=-\epsilon_{kil}\hat{G}_{k\!j,l}
=\epsilon_{ikl}\hat{G}_{k\!j,l}
\) 
from \eqref{equ:FielTenSecOrdMalDef}${}_{1}$, corresponding to 
\(
-\lbrack
\mathrm{curl}_{4}\hat{\bm{G}}
\rbrack_{i\!j}
\) 
via \eqref{equ:CurMalCor}${}_{1}$. The form for \(\hat{\alpha}_{i\!j}\) 
in their Equation (22) 
disagrees with this and in fact is mathematically incorrect. Later work 
employing the approach of \cite{Hartley2005a} corrected this; for example, 
\citet{Hirel2015} works with 
\(
\hat{\alpha}_{i\!j}
=-\hat{G}_{ki,l}\epsilon_{klj}
=\hat{G}_{ki,l}\epsilon_{kjl}
=\lbrack\mathrm{curl}_{3}\hat{\bm{G}}\rbrack_{i\!j}
\) 
from \eqref{equ:ComCarCurTenOrdSecTwo}${}_{2}$ 
in the software package Atomsk (https://atomsk.univ-lille.fr). 

\end{appendix}

\end{document}